# Spectral Duality for Planar Billiards

*J.-P. Eckmann*[1,2] *and C.-A. Pillet*[1]

[1]Dépt. de Physique Théorique, Université de Genève, CH-1211 Genève 4, Switzerland
[2]Section de Mathématiques, Université de Genève, CH-1211 Genève 4, Switzerland

**Abstract.** For a bounded open domain $\Omega$ with connected complement in $\mathbf{R}^2$ and piecewise smooth boundary, we consider the Dirichlet Laplacian $-\Delta_\Omega$ on $\Omega$ and the S-matrix on the complement $\Omega^c$. We show that the on-shell S-matrices $\mathbf{S}_k$ have eigenvalues converging to 1 as $k \uparrow k_0$ exactly when $-\Delta_\Omega$ has an eigenvalue at energy $k_0^2$. This includes multiplicities, and proves a weak form of "transparency" at $k = k_0$. We also show that stronger forms of transparency, such as $\mathbf{S}_{k_0}$ having an eigenvalue 1 are not expected to hold in general.

In this paper, we consider a simply connected bounded domain $\Omega$ in $\mathbf{R}^2$, with piecewise smooth boundary $\Gamma = \partial\Omega$. We establish a correspondence between the eigenvalues of the Dirichlet Laplacian in $\Omega$, and the S-matrix (also with Dirichlet condition) for the exterior domain $\Omega^c$. In its crudest form, this relation says that *$k^2 = E$ is an eigenvalue of the "inside problem" if and only if the on-shell S-matrix has an eigenvalue 1 at that energy.* This relation has been conjectured in [DS] and subsequently studied numerically in [DS1], [DS2], with an excellent agreement. Furthermore, in the semi-classical limit, this relation leads to a new derivation [DS] of the Gutzwiller trace formula [Gu]. For an exposition of this and related problems in quantum billiards, we refer the reader to [S]. One can reformulate the conjecture to say that the obstacle is transparent for a carefully selected wave, whenever one scatters at an energy which is equal to an eigenenergy of the Dirichlet Laplacian. The basic idea of the conjecture is that the scattering wave function and the inside eigenfunction are simply one and the same function which happens to vanish on the boundary $\Gamma$. It is an easy exercise to check that this conjecture holds for a *one*-dimensional billiard, i.e., for a Laplacian on an interval with zero boundary conditions and on its complement [F].

However, in 2 or more dimensions, this "inside-outside duality" (or "spectral duality") does not hold in the exact form given above, but only in a slightly weaker sense. In order to formulate our result, we will need some machinery which is developed below, but we can already describe the main flavor of the statement in an informal way:

1. If the S-matrix has an eigenvalue 1 at some energy $E$, then this energy is an eigenenergy of the inside problem. In this case, the interior eigenfunction can be continued to a *bounded* solution of the Helmholtz equation in the full plane.
2. If $E$ is an eigenvalue of the inside problem, then for $E'$ close to, but below, $E$, the S-matrix has an eigenvalue $e^{-2i\vartheta(E')}$, with $0 < \vartheta(E') < \pi$. As $E' \uparrow E$, the angle $\vartheta(E')$ reaches $\pi$ from below. Conversely, if $\vartheta(E') \uparrow \pi$ as $E' \uparrow E$, then $E$ is an eigenvalue of the inside problem.

The formulation given above may seem overly cautious, but the statement covers the (probably generic) case when the *eigenfunction* of the S-matrix does not exist for $E' = E$.

Still, for all nearby $E' < E$ there will be eigenfunction, and the corresponding eigenvalues converge to 1. We will give examples where the S-matrix does not have an eigenfunction for energies corresponding to the inside problem, because the inside eigenfunction can simply not be extended to the full plane $\mathbf{R}^2$. In [B], an example of a domain $\Omega$ is given for which the extension of the eigenfunction is unbounded. This provides another class of domains for which the S-matrix does not have an eigenvalue 1 on the energy shell $E$.

The basic idea underlying the analysis is the application of potential theory to this problem, combined with some functional analysis. The potential theory aspects are exposed for example in [R] or in [KR], but for the convenience of the reader, the relevant features of this theory will be explained here. We will connect the scattering theory and the eigenvalue problem by expressing both the resolvent of the inner Laplacian and the scattering matrix of the outer problem in terms of the single layer potential on the common boundary $\Gamma$. We then characterize the spectrum of the S-matrix by a variational formula.

The paper is organized as follows. In Sect.1 we define the S-matrix and we formulate the results (Main Theorem). We also give examples for which the S-matrix does not have an eigenvalue 1 at $E$. In Sect.2 we present the potential theory aspects of the problem. They involve in particular the Green's function, restricted to the boundary of the billiard. We also define a modified S-matrix, which acts on the boundary, and which has the same spectrum as the conventional S-matrix. This is useful for applications [DS1], [DS2]. In Sect.3 we prove that the boundary restriction operator is Fredholm. It is here that the restrictions on the shape of the domain are crucial. In Sect.4 we establish a resolvent formula, and express the S-matrix in terms of the boundary restriction operator. Equipped with this information, we characterize in Sect.5 the eigenvalues of the S-matrix as the solution of a variational problem, establishing the spectral duality.

In a subsequent paper with U. Smilansky and I. Ussishkin [EPSU], we plan to give numerical examples of the precise meaning of the Main Theorem.

## 1. Definition of the S-matrix and statement of the results

In this paper, we shall give proofs of the spectral duality for piecewise smooth bounded domains $\Omega$:

**Definition.** A *standard domain* $\Omega$ is a simply connected bounded domain in $\mathbf{R}^2$ whose boundary $\Gamma = \partial \Omega$ is piecewise $\mathcal{C}^2$. By this we mean that $\Gamma$ has a *finite* number of differentiable pieces. Furthermore, we require the angles at the corners to be bounded away from 0 and $2\pi$. Finally, we always assume $\Omega$ is non-empty.

**Remarks.**
1. We do **not** assume that $\Omega$ is convex, and the difficulties with the spectral duality are not related to convexity.
2. We note the slightly astonishing fact that the proofs given in this paper generalize with only notational differences to the case of a *finite union of standard domains*, replacing $\Gamma$ by $\cup_{j=1}^{N} \Gamma_j$. But we really need that $\mathbf{R}^2 \setminus \Omega$ is connected.

**Notation.** We denote by $\Delta_\Omega$ the Laplacian in $\Omega$ with Dirichlet boundary conditions on $\Gamma$, and by $\sigma(\Delta_\Omega)$ its spectrum. We let $\Omega^c$ denote the exterior of the billiard and $\Delta_{\Omega^c}$ the corresponding Dirichlet Laplacian.

We next define the quantum-mechanical S-matrix. For a "free" Hamiltonian $H_0$ and an interacting Hamiltonian $H$, it is given by the formula

$$\mathbf{S} = \operatorname*{s-lim}_{\varepsilon \downarrow 0} \varepsilon \int_0^\infty dt\, e^{-\varepsilon t} e^{iH_0 t} e^{-2iHt} e^{iH_0 t} , \tag{1.1}$$

where s-lim denotes the strong limit. In our case, $-H_0 = \Delta$ and $-H = \Delta_\Omega \oplus \Delta_{\Omega^c}$. By energy conservation $\mathbf{S}$ can be decomposed as a sum over the on-shell S-matrices $\mathbf{S}_k$ which act on $L^2$ of the energy shell $F_k = \{p \in \mathbf{R}^2 \mid p^2 = k^2\}$. A detailed formula will be given in the next section. The following lemma describes the eigenvalues of the on-shell S-matrix:

**Lemma 1.1.** *Let $\Omega$ be a standard domain, and let $k > 0$. Then the operator $\mathbf{S}_k$ is unitary with spectrum on the unit circle. It consists of eigenvalues of finite multiplicity, accumulating only at $1$. Furthermore, they accumulate there only from below.*

**Remark.** Similar statements can be found in [Y1, Y2, JK].

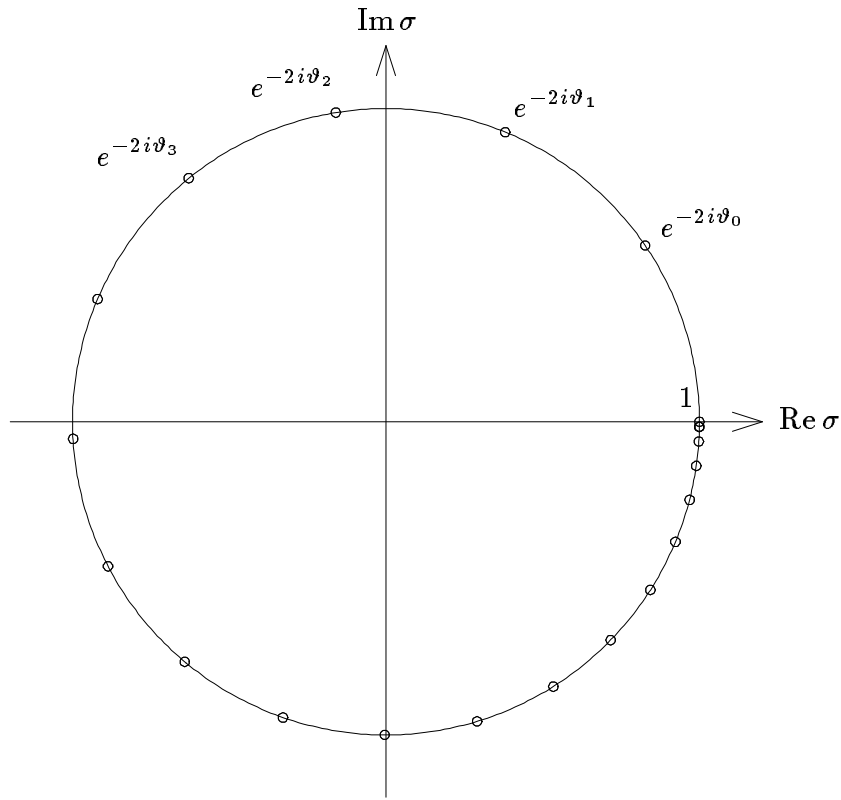

**Fig. 1**: The qualitative aspect of the spectrum of the S-matrix. Note that eigenvalues accumulate at 1 only from below.

This will be shown in Sect.4. The spectrum is illustrated qualitatively in Fig. 1. We next fix $k > 0$. By the lemma, we can write the eigenvalues of $\mathbf{S}_k$ as $e^{-2i\vartheta_j(k)}$, and we order these *scattering phases* $\vartheta_j$, $j = 0, 1, \ldots$, by

$$\pi > \vartheta_0 \geq \vartheta_1 \geq \vartheta_2 \geq \cdots \geq 0 . \tag{1.2}$$

While 0 is always an accumulation point of the $\vartheta_j$, it might not correspond to an eigenvalue. We can now formulate the spectral duality result:

**Main Theorem.** *Let $\Omega$ be a standard domain. Then the following two statements are equivalent:*
1. *The Laplacian $-\Delta_\Omega$ has an $M$-fold degenerate eigenvalue $k_0^2$.*
2. *As $k \uparrow k_0$, exactly $M$ eigenphases $\vartheta_j(k)$ of the S-matrix $\mathbf{S}_k$ converge to $\pi$ from below.*

If $\mathbf{S}_k$ has an eigenvalue 1, then we can state the simpler

**Theorem 1.3.** *Let $\Omega$ be a standard domain. If the operator $\mathbf{S}_k$, $k > 0$, has an eigenvalue 1 of multiplicity $M$ with eigenvectors in $L^2(F_k)$, then $-\Delta_\Omega$ has an eigenvalue $k^2$ of multiplicity at least $M$. Furthermore, the corresponding Dirichlet eigenfunctions can be extended to bounded solutions of the Helmholtz equation in all of $\mathbf{R}^2$.*

**Remarks.**
1. The proofs will be given in Sect.5, by using a variational principle. Our results deal with the behavior of the eigenvalues of $\mathbf{S}_k$ for $k < k_0$. Although these eigenvalues simply cross 1 for scattering from a circle, numerical studies [EPSU] seem to indicate that for a general domain, non-analytic behavior at $k = k_0$ is to be expected.
2. We present the theory only for the case of Dirichlet boundary conditions. The extension to other conditions should be rather straightforward. Also, the study of this paper is restricted to 2 dimensional domains. We conjecture that the results extend to higher dimensions, but this needs a definition of standard domains in higher dimensions for which the methods of Sect.3 are applicable.
3. For a discussion of some numerical aspects, see the end of Sect.2.

As mentioned in the introduction, one could think that spectral duality holds in one of the following stronger forms: The inside eigenvalues are in one-to-one correspondence with those energies where the on-shell S-matrix has an eigenvalue 1, or, a specific scattering wave extends to an eigenfunction of $\Delta_\Omega$. It has been noticed earlier that such stronger forms hold when $\Omega$ is a disc, an ellipse, or a rectangle [DS1, DS2]. We now show that there are domains where for some (or all) $k^2 \in \sigma(-\Delta_\Omega)$, the operator $\mathbf{S}_k$ does *not* have an eigenvalue 1, so that neither of the stronger forms of spectral duality hold.

**Example 1. The cake.** Consider the domain

$$\Omega = \{(r, \varphi) : 0 < r < 1, |\varphi| < \pi/3\} , \tag{1.3}$$

written in polar coordinates. For this domain, $\sigma(-\Delta_\Omega) = \{k_{\ell,n}^2, \ell, n = 1, 2, \ldots\}$, where $k_{\ell,n}$ is the $n^{\text{th}}$ nontrivial zero of the Bessel function $J_{3\ell/2}(x)$ and the corresponding eigenfunctions are

$$\psi_{\ell,n}(r, \varphi) = J_{3\ell/2}(k_{\ell,n} r) \cdot \cos(3\ell\varphi/2) .$$

When $\ell$ is odd, these functions do not extend to $\mathbf{R}^2$ because they are not $2\pi$-periodic in $\varphi$, and hence $\mathbf{S}_{k_{\ell,n}}$ cannot have an eigenvalue 1 by Theorem 1.3. Note that the eigenfunctions have their branch points on the boundary of the domain. In Examples 2 and 3 below the singularity lies outside the domain.

**Example 1a. The irrational cake.** Consider the domain

$$\Omega = \{(r,\varphi) \ : \ 0 < r < 1, \ |\varphi| < \nu\pi\} \,, \tag{1.4}$$

where $\nu$ is irrational. Then none of the inside eigenfunctions (which are still explicitly known), can be continued outside.

**Example 2. Smooth boundary.** We define

$$C(\rho,\psi) = J_\nu(k_\nu \rho) \cdot \cos(\nu\psi) \,,$$

where $k_\nu$ is the first nontrivial zero of $J_\nu$. In the sequel, we take $\nu = 3/2$, but any other non-integer $\nu$ would be just as good. Note that $\rho = 0$ is a branch point of the cake function $C(\rho,\psi)$. We construct a new function, fixing $p \in \mathbf{Z}^+$:

$$R(r,\varphi) = \sum_{j=0}^{p-1} C(\rho_j, \psi_j) \,. \tag{1.5}$$

Here, we fix $t > 0$ and define $\varphi_j = \varphi + 2\pi j/p$,

$$\begin{aligned} x_j &= t + r\cos(\varphi_j) \,, \\ y_j &= r\sin(\varphi_j) \,. \end{aligned}$$

Finally,
$$\begin{aligned} \rho_j \cos(\psi_j) &= x_j \,, \\ \rho_j \sin(\psi_j) &= y_j \,. \end{aligned}$$

Note that $\rho_j = 0$ if $r = t$ and $\varphi_j = \pm\pi$, i.e., if $\varphi = (2j/p - 1)\pi$. We define the curve $\Gamma$ as the zero level set of $R$ near the origin, see Fig. 2.

Then, $R$ is a Dirichlet eigenfunction with eigenvalue $k_\nu^2$ for the corresponding $\Omega$, which is smooth and convex, but $R$ has branch points strictly outside $\Omega$.

**Example 3. Smooth boundary and a dense set of singularities.** One can construct an example with a convex boundary and a set of singularities which are dense on a circle. Let $R(r,\varphi)$ be the function defined in Example 2 and let $\Gamma$ be the zero level curve of this function. Fix a large radius $r_0$, and enumerate the rational points on this circle, with angles $\alpha_n, n = 1, \ldots$ . Define

$$F(r,\varphi) = \sum_{n=0}^\infty \frac{C(\rho_n, \psi_n)}{n!n!} \,,$$

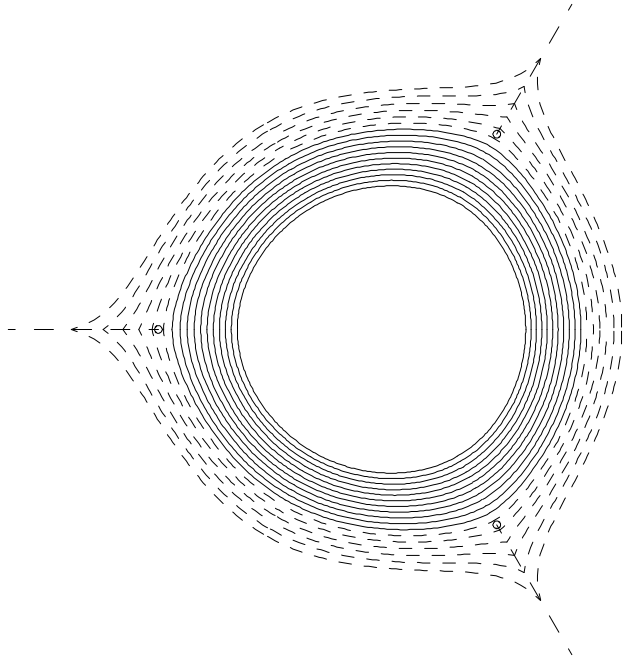

**Fig. 2**: The level curves for the function $R$ defined in Eq.(1.5), for $p = 3$. We have chosen $t = 0.6$. Solid lines correspond to positive values of $R$, dashed lines to negative ones, with a level spacing of 0.06. The three branch points, with their cuts, are marked by circles. The outermost solid line is the boundary of a domain $\Omega$ with an interior eigenfunction which cannot be continued into the complement $\Omega^c$.

where
$$x_n = r_0 - r\cos(\varphi - \alpha_n),$$
$$y_n = r\sin(\varphi - \alpha_n),$$

and
$$\rho_n \cos(\psi_n) = x_n,$$
$$\rho_n \sin(\psi_n) = y_n.$$

Note that we have $\rho_n = 0$ if $r = r_0$ and $\varphi = \alpha_n$. Thus, $F(r,\varphi)$ has branch points at all points $(r_0, \alpha_n)$, since the sum converges by the choice of our very large denominator. In fact, on every compact set, $|F|$ is uniformly bounded, and it is analytic for $r < r_0$. For all $\varepsilon$, the function

$$K(r,\varphi) = R(r,\varphi) + \varepsilon F(r,\varphi)$$

has singularities at the three points determined by $R$ *and* on the rational points of the circle of radius $r_0$. Furthermore, when $\varepsilon$ is very small the level zero curve $\Gamma_\varepsilon$ of $K$ is very close to $\Gamma$, and since $F$ is analytic near $\Gamma$, the curve remains strictly convex if $\varepsilon > 0$ is sufficiently small. Let $\Omega_\varepsilon$ be the domain whose boundary is $\Gamma_\varepsilon$. Then $K$ is an eigenfunction of $-\Delta_{\Omega_\varepsilon}$ with eigenvalue $k^2_{3/2}$. It cannot be continued beyond the circle of radius $r_0$.

## 2. Potential theory

In this section we present notions from potential theory which will be used throughout. This allows us to formulate the strategy of the proof, as well as some results connected to numerical calculations. After introducing some function spaces, we will define the restriction $\gamma$ of a function to $\Gamma$, and the "single layer potentials" $\mathbf{G}_k^\pm$ and their very important "boundary restriction" operators $\mathbf{A}_k$.

The natural spaces on which we consider the problem are $L^2$ spaces, and Sobolev spaces. In order to define these spaces, we introduce a new system of coordinates, the arclength along $\Gamma$. We call the corresponding variables $s$, $s'$. Thus, $s$ varies in $I_L = [-L/2, L/2]$, and there is a periodic map $x : I_L \to \mathbf{R}^2$ which maps $I_L$ onto the curve $\Gamma \subset \mathbf{R}^2$. The space $\mathcal{H}_\Gamma$ is the space of $L^2$ functions on the boundary $\Gamma$, with the measure $ds$. The Sobolev spaces $\mathcal{H}_\Gamma^\beta$ are defined in the usual way: Denoting by $\partial_s$ the derivative with periodic boundary conditions on $I_L$ and setting $\Lambda = (1 + (i\partial_s)^2)^{1/2}$, we define, for $\beta \geq 0$,

$$\mathcal{H}_\Gamma^\beta = \{u \in \mathcal{H}_\Gamma : \Lambda^\beta u \in \mathcal{H}_\Gamma\} .$$

To simplify notation, we write

$$\int_{I_L} ds\, \psi(x(s)) = \int_\Gamma d\sigma(z)\, \psi(z) .$$

**Notation.** When no confusion is possible, we write $k$ instead of $|k|$, for $k \in \mathbf{R}^2$, and similarly for other coordinates. If $k^2 \neq 0$, then *we always tacitly assume that $k > 0$.*

**Notation.** The letters $u, v, \ldots$ denote functions on the boundary $\Gamma$, the letters $\psi, \varphi, \ldots$ denote functions in $\mathbf{R}^2$ (or in $\Omega, \Omega^c$), and $\chi$ denotes a function (of $p$) on the energy shell $F_k$.

**Definition and properties of $\gamma$ and $\gamma^*$.** The restriction to $\Gamma$ is given by

$$(\gamma\psi)(z) = \psi(z) , \quad \text{when } z \in \Gamma ,$$
$$(\gamma^* u, \psi)_{L^2(\mathbf{R}^2)} = \int_\Gamma d\sigma(z)\, \bar{u}(z)\psi(z) .$$

For the boundary $\Gamma$ of a standard domain $\Omega$ one has the following classical results [Ne] for $\gamma$:

$$\begin{aligned}
\gamma &: H_{\text{loc}}^\beta(\mathbf{R}^2) \to \mathcal{H}_\Gamma , \quad \text{for all } \beta > \tfrac{1}{2} , \\
\gamma^* &: \mathcal{H}_\Gamma \to H_{\text{comp}}^{-\beta}(\mathbf{R}^2) , \quad \text{for all } \beta > \tfrac{1}{2} , \\
\gamma &: H_{\text{loc}}^1(\mathbf{R}^2) \to \mathcal{H}_\Gamma^{1/2} , \\
\gamma^* &: \mathcal{H}_\Gamma^{-1/2} \to H_{\text{comp}}^{-1}(\mathbf{R}^2) ,
\end{aligned} \quad (2.1)$$

and,

$$\ker(\gamma^*) = \{0\} . \tag{2.2}$$

Here, $H_{\text{comp}}$ is the subspace of functions with compact support in $H$, and $H_{\text{loc}}$ are the function which are locally in $H$, see [H].

**Definition and properties of $\mathbf{G}_k^\pm$ and $\mathbf{A}_k$.** Here, we introduce the central objects, the "single layer potential" **G** and the "boundary restriction" **A**. We denote by $G$ the free space Green's function in $\mathbf{R}^2$:

$$G(\zeta) = (-\Delta - \zeta)^{-1}, \tag{2.3}$$

and, for fixed energy $E = k^2, k > 0$,

$$G_k^\pm = G(k^2 \pm i0). \tag{2.4}$$

For this operator, one has [H]

$$G_k^\pm : H_{\text{comp}}^{-\beta}(\mathbf{R}^2) \to H_{\text{loc}}^{-\beta+2}(\mathbf{R}^2), \quad \text{for all } \beta. \tag{2.5}$$

Then we define the single layer potentials by

$$(\mathbf{G}_k^\pm u)(x) = \int_\Gamma d\sigma(z)\, G_k^\pm(x - z)u(z). \tag{2.6}$$

In other words,

$$\mathbf{G}_k^\pm = G_k^\pm \gamma^*. \tag{2.7}$$

Combining (2.1), (2.5), and (2.7), we see that

$$\mathbf{G}_k^\pm : \mathcal{H}_\Gamma \to H_{\text{loc}}^\beta(\mathbf{R}^2), \quad \text{for all } \beta < \tfrac{3}{2}. \tag{2.8}$$

This means in particular, that **G** maps to continuous functions. Furthermore, from $(-\Delta - k^2)\mathbf{G}_k^\pm u = \gamma^* u$, we see that $\mathbf{G}_k^\pm u$ solves the Helmholtz equation in $\mathbf{R}^2 \setminus \Gamma$. By Eq.(2.2) it follows that

$$\ker(\mathbf{G}_k^\pm) = \{0\}. \tag{2.9}$$

Coming back to Eq.(2.8), we can define the "boundary restriction" operator

$$(\mathbf{A}_k u)(z) = (\mathbf{G}_k^+ u)(z), \tag{2.10}$$

for $z \in \Gamma$. In other words, $\mathbf{A}_k = \gamma G_k^+ \gamma^*$. It follows that

$$\mathbf{A}_k : \mathcal{H}_\Gamma \to \mathcal{H}_\Gamma. \tag{2.11}$$

It follows at once from the definition that

$$\mathbf{A}_k^* = \bar{\mathbf{A}}_k = \gamma G_k^- \gamma^*, \tag{2.12}$$

where $\mathbf{A}^*$ is the adjoint and $\bar{\mathbf{A}}$ is the complex conjugate. In Sect.3 we will show that for standard domains, one has the stronger result: $\mathbf{A}_k : \mathcal{H}_\Gamma \to \mathcal{H}_\Gamma^1$.

It will be important to consider the decomposition of $\mathbf{A}_k$ into its real and imaginary parts:

$$\mathbf{A}_k = \mathbf{Y}_k + i\mathbf{J}_k, \tag{2.13}$$

where $\mathbf{Y}_k$ and $\mathbf{J}_k$ are real, self-adjoint operators. This notation reflects the decomposition of $G$ into Hankel and Bessel functions:

$$G_k^{\pm}(x) = (i/4)H_0^{\pm}(k|x|) = (i/4)J_0(k|x|) \mp (1/4)Y_0(k|x|).$$

Note that $J_0$ is entire analytic, and $Y_0$ has a logarithmic singularity at 0.

**Strategy of proof, and numerical aspects.** Our proof of the Main Theorem will be based on a number of identities which we now list without specifying domains of applicability. Starting with the operator $\mathbf{J}_k$, one can write it as

$$\mathbf{J}_k = \operatorname{Im} \mathbf{A}_k = \pi \mathcal{L}_k^* \mathcal{L}_k, \tag{2.14}$$

where $\mathcal{L}_k$ maps functions on the boundary $\Gamma$ to functions on the energy shell $F_k$. With these notations we have two important identities:
1. The on-shell S-matrix $\mathbf{S}_k$ is given by

$$\mathbf{S}_k = 1 - 2\pi i \mathcal{L}_k \mathbf{A}_k^{-1} \mathcal{L}_k^*. \tag{2.15}$$

2. The eigenenergies of $-\Delta_\Omega$ are exactly those $k^2$ for which $\mathbf{A}_k u = 0$ has non-trivial solutions. (This is a well-known result from potential theory.)

Using the intimate relations between $\mathcal{L}_k$ and $\mathbf{A}_k$ on can define a modified S-matrix which acts on functions on the boundary alone, which is given by

$$\widetilde{\mathbf{S}}_k = \mathbf{A}_k^* \mathbf{A}_k^{-1}. \tag{2.16}$$

This operator has the same spectrum as $\mathbf{S}_k$ and seems to be useful for doing numerics [DS1, DS2].

## 3. The Fredholm property of the boundary restriction operator $\mathbf{A}_k$

In this section we study the operator $\mathbf{A}_k$ on the Sobolev spaces $\mathcal{H}_\Gamma^\beta$. We shall use mostly the coordinates $s \in I_L$, and the map $x : I_L \to \Gamma \subset \mathbf{R}^2$ defined in Sect.2. The operator $\mathbf{A}_k$ has then an integral kernel $\mathbf{A}_k(s, s')$ (as a map from $L^2(I_L)$ to itself), given by

$$\mathbf{A}_k(s, s') = G_k^+(x(s), x(s')).$$

Recall the decomposition $\mathbf{A}_k = \mathbf{Y}_k + i\mathbf{J}_k$. The main result of this section is:

**Theorem 3.1.** *Let $\Omega$ be a standard domain and $k > 0$. Let $\Lambda = \left(1 + (i\partial_s)^2\right)^{1/2}$. Then, for all $\beta \in [0, 1]$, the operator*

$$\Lambda^{1-\beta} \mathbf{A}_k \Lambda^\beta \tag{3.1}$$

is bounded and Fredholm on $\mathcal{H}_\Gamma$ and has index 0. Furthermore, one has, for $\beta = 1/2$, the representations

$$\Lambda^{1/2}\mathbf{Y}_k\Lambda^{1/2} = \tfrac{1}{2}\mathbf{1} + \mathbf{B} + \mathcal{K}_{k,1}, \tag{3.2}$$

$$\Lambda^{1/2}\mathbf{J}_k\Lambda^{1/2} = \mathcal{K}_{k,2} \geq 0. \tag{3.3}$$

The operator $\mathbf{B}$ is independent of $k$, self-adjoint, and bounded, $\|\mathbf{B}\| = r < \tfrac{1}{2}$. Finally, $\mathcal{K}_{k,1}$ and $\mathcal{K}_{k,2}$ are compact and they are analytic in $\{k \mid k \in \mathbf{C} \setminus 0\}$.

**Corollary 3.2.** *Let $\Omega$ be a standard domain and let $k > 0$. Then $\mathbf{Y}_k = \operatorname{Re}\mathbf{A}_k$ has a finite number of negative eigenvalues.*

**Corollary 3.3.** *Let $\Omega$ be a standard domain and let $k > 0$. Then*

$$\ker\left(\mathbf{A}_k|_{\mathcal{H}_\Gamma^{-\beta}}\right) = \ker\left(\mathbf{A}_k|_{\mathcal{H}_\Gamma}\right), \tag{3.4}$$

*for all $\beta \in [0,1]$.*

**Remarks.**
1. One can express Eq.(3.1) in terms of the spaces $\mathcal{H}_\Gamma^\beta$: $\mathbf{A}_k$ is a map

$$\mathbf{A}_k|_{\mathcal{H}_\Gamma^{-\beta}} : \mathcal{H}_\Gamma^{-\beta} \to \mathcal{H}_\Gamma^{1-\beta}. \tag{3.5}$$

Similarly, Eq.(3.4) says that every function in the kernel of $\mathbf{A}_k|_{\mathcal{H}_\Gamma^{-\beta}}$ is in the more regular space $\mathcal{H}_\Gamma$.

2. If $\mathbf{A}_k|_{\mathcal{H}_\Gamma^{-\beta}}$ is Fredholm, this means that $\mathbf{A}_k^{-1}$ is bounded from $\mathcal{H}_\Gamma^{1-\beta}$ to $\mathcal{H}_\Gamma^{-\beta}$, whenever $\ker(\mathbf{A}_k) = \{0\}$. It is this property which is used throughout the paper. In fact, the proof of Theorem 3.1 will give a rather detailed description of the essential spectrum of $\Lambda\mathbf{A}$.

3. The proof of Theorem 3.1 is straightforward, but a little long, and this is due to the class of domains we want to handle. For example, if $\Omega$ has a *smooth* boundary, then the corresponding result is known, and is spelled out in [R]. On the other hand, even in the case we consider, there is a large body of results describing the boundary behavior of eigenfunctions of $-\Delta_\Omega$. In particular, the lectures of Agmon [A], as well as a lot of subsequent literature (see e.g., [GT, Ne]), deal with domains which have the "uniform exterior cone property" and our definition of standard domain is a slightly stronger version of this property, adapted to the case of 2 dimensions. (The strengthening is that we allow only for a finite number of corners.) Although the literature contains detailed information about the boundary behavior, we have not been able to extract the Theorem 3.1 from it. Therefore we give here a self-contained proof of Theorem 3.1.

4. It will follow from the proof that all the bounds are also valid upon replacing $k^2 + i0$ by an arbitrary complex number $z \neq 0$.

**Proof of Theorem 3.1.** The proof will take up most of this section, and its details are independent of the other developments of this paper. We omit the index $k$ in the sequel. We begin by showing that $\mathbf{A} : \mathcal{H}_\Gamma \to \mathcal{H}_\Gamma^1$ is Fredholm, and we will extend this later to arbitrary $\beta$. More precisely, we differentiate and show that

$$|i\partial_s|\mathbf{A} \; : \; L^2(I_L) \to L^2(I_L) \tag{3.6}$$

is Fredholm.

Since we are interested only in the essential spectrum of $|i\partial_s|\mathbf{A}$, it is useful to introduce the notation $\approx$ for equivalence up to compact operators. Note that any piece $P$ of $\mathbf{A}$ for which $i\partial_s P(s,s')$ is compact can be eliminated [K]. Indeed, if $i\partial_s P(s,s')$ is compact, then $|i\partial_s|P(s,s')$ is compact as well, since $|i\partial_s| = \text{sign}(i\partial_s) \cdot i\partial_s$ and $\text{sign}(i\partial_s)$ is a bounded operator.

We start the proof by noting that the Green's function for the Helmholtz operator is the Hankel function [AS]:

$$\mathbf{A}_k(s,s') \;=\; G_k^+(x(s), x(s')) \;=\; \frac{i}{4} H_0^+(k|x(s) - x(s')|) \, .$$

The known singularity of $H_0^+$ leads to the representation

$$\mathbf{A}(s,s') \;=\; -\frac{1}{2\pi}\log|x(s) - x(s')| + V(s,s') \;\equiv\; \mathbf{A}^{(0)}(s,s') + \mathbf{A}^{(1)}(s,s') \, .$$

The function $\mathbf{A}^{(1)}$ is the sum of terms of the form $(x(s) - x(s'))^{2n-2}$ and $\log|x(s) - x(s')| \cdot (x(s) - x(s'))^{2n}$, $n \geq 1$, [AS, 9.1.12–13]. Since $\Gamma$ is bounded, $i\partial_s \mathbf{A}^{(1)}(s,s')$ is bounded as well, and hence $\mathbf{A} \approx \mathbf{A}^{(0)}$. It suffices therefore to analyze $\mathbf{A}^{(0)}$. We write it as

$$\begin{aligned}\mathbf{A}^{(0)}(s,s') &= -\frac{1}{2\pi}\log|x(s) - x(s')| \\ &= -\frac{1}{2\pi}\log|\sin(\frac{\pi}{L}(s-s'))| - \frac{1}{2\pi}\log\left|\frac{x(s)-x(s')}{\sin(\pi(s-s')/L)}\right| \equiv \mathbf{A}^{(2)} + \mathbf{A}^{(3)} \, .\end{aligned}$$

We want to consider first the term $\mathbf{A}^{(2)}$ which will be identified below as the main term. We start with some useful identities:

**Lemma 3.4.** *One has the following identities for the integral kernels:*

$$|i\partial_s|^{-1}(s,s') = -\frac{1}{2\pi}\log\bigl(4\sin^2(\pi(s-s')/L)\bigr) \, , \tag{3.7}$$

$$\text{sign}(i\partial_s)(s,s') = \frac{i}{L}\cot(\pi(s-s')/L) \, . \tag{3.8}$$

**Proof.** We consider on $L^2(I_L)$ the generator of translations $i\partial_s$ with periodic boundary conditions. An orthonormal eigenbasis is given by the functions $\varphi_\ell(s) \equiv L^{-1/2}e^{i2\pi\ell s/L}$, for which $i\partial_s \varphi_\ell = -\ell \cdot 2\pi L^{-1} \varphi_\ell$. Thus, the operator $|i\partial_s|$ is invertible on the orthogonal complement $L_0^2(I_L) \subset L^2(I_L)$ of the constant functions. The integral kernel of the inverse is then

$$|i\partial_s|^{-1}(s,s') = \sum_{\ell \neq 0} \frac{L}{2\pi|\ell|} \varphi_\ell(s)\bar\varphi_\ell(s') = \sum_{\ell \neq 0} \frac{1}{2\pi|\ell|} e^{i2\ell\pi(s-s')/L}$$

$$= \frac{1}{2\pi} \sum_{\ell=1}^{\infty} \frac{1}{\ell}(z^\ell + \bar z^\ell)\Big|_{z=e^{i2\pi(s-s')/L}} .$$

The sum is readily evaluated by first considering $|z| < 1$ and then taking the limit and one obtains

$$\frac{1}{2\pi} \sum_{\ell=1}^{\infty} \frac{1}{\ell}(z^\ell + \bar z^\ell) = -\frac{1}{2\pi}\big(\log(1-z) + \log(1-\bar z)\big) = -\frac{1}{2\pi}\log(1 + |z|^2 - 2\operatorname{Re} z) .$$

When $z = e^{i\vartheta}$, this leads to

$$\frac{1}{2\pi}\sum_{\ell \neq 0} \frac{1}{|\ell|} e^{i\vartheta \ell} = -\frac{1}{2\pi}\log\big(2(1-\cos\vartheta)\big) = -\frac{1}{2\pi}\log\big(4\sin^2(\vartheta/2)\big) . \tag{3.9}$$

From this, we find Eq.(3.7). Upon differentiating Eq.(3.9) w.r.t. $\vartheta$ we obtain in addition (3.8). The proof of Lemma 3.4 is complete. $\square$

We continue the proof of Theorem 3.1. By Lemma 3.4, we find

$$\mathbf{A}^{(2)}(s,s') = -(2\pi)^{-1}\log|\sin(\pi(s-s')/L)| = \tfrac{1}{2}|i\partial_s|^{-1}(s,s') + (2\pi)^{-1}\log 4 .$$

Therefore,
$$|i\partial_s|\mathbf{A} = \tfrac{1}{2}(\mathbf{1} - P_{\text{const}}) + |i\partial_s|\mathbf{A}^{(3)} \approx \tfrac{1}{2}\mathbf{1} + |i\partial_s|\mathbf{A}^{(3)} , \tag{3.10}$$

where $P_{\text{const}}$ is the projection onto constant functions.

**Remark.** Although the study of $|i\partial_s|$ is more complicated than that of $i\partial_s$, we have preferred it because it leads to the appearance of the operator $\tfrac{1}{2}\mathbf{1}$ in Eq.(3.10).

We next study $\mathbf{A}^{(3)}$. Not all of its contributions are negligible, and in fact the corners play an important rôle. In order to isolate their contribution, we need a variety of cutoffs. We use a cutoff function $h \in \mathcal{C}^\infty$, which is symmetric, of compact support and equal to 1 near the origin. We start by isolating the irrelevant parts of $\mathbf{A}^{(3)}$. We have the

**Lemma 3.5.** *If $h$ has sufficiently small support, then*

$$i\partial_s\left(1 - h\big(\sin(\frac{\pi(s-s')}{L})\big)\right)\mathbf{A}^{(3)}(s,s') \approx 0 .$$

**Proof.** By the chain rule, we find, with $\zeta(s-s') = \sin(\pi(s-s')/L)$,

$$\partial_s(1-h)\mathbf{A}^{(3)} = -\frac{\pi}{L}\cos(\pi(s-s')/L)h'(\zeta)\mathbf{A}^{(3)} + (1-h(\zeta))\partial_s\mathbf{A}^{(3)}(s,s') .$$

Note that both $h'(\zeta)$ and $1 - h(\zeta)$ vanish near the diagonal $s = s'$, and that $\mathbf{A}^{(3)}$ and $\partial_s \mathbf{A}^{(3)}$ are bounded outside any open neighborhood of the diagonal. Therefore, the differentiability of $x(s)$ away from the corners implies that $i\partial_s(1 - h(\zeta))\mathbf{A}^{(3)}$ is Hilbert-Schmidt, and hence compact. The proof of Lemma 3.5 is complete. □

Thus, we are led to study $\mathbf{A}^{(4)}(s,s') \equiv h(\zeta(s-s'))\mathbf{A}^{(3)}(s,s')$. We let $s_j, j = 1, \ldots, N$, be the position of the $j^{\text{th}}$ corner. We assume that the support of $h$ is so small that the $h(\zeta(s-s_j))$ have disjoint supports. We next consider $\mathbf{A}^{(4)}$ away from the corners, which leads to another irrelevant piece.

**Lemma 3.6.** *If $h$ has sufficiently small support, then*

$$i\partial_s \left(1 - \sum_{j=1}^{N} h(\zeta(s' - s_j))\right) \mathbf{A}^{(4)}(s,s') \approx 0 .$$

**Proof.** By the chain rule, we have

$$\partial_s (1 - \sum h)\mathbf{A}^{(4)} = (1 - \sum h) \cdot \left(\frac{\pi}{L}\cos(\pi(s-s')/L)h'(\zeta)\mathbf{A}^{(3)} + h(\zeta)\partial_s\mathbf{A}^{(3)}\right) .$$

We have already seen above that the first term leads to a compact operator. The second term has support near the diagonal, but away from the corners. *We are now using that $x(s)$ is $\mathcal{C}^2$ away from the corners.* This implies that

$$\partial_s \mathbf{A}^{(3)}(s,s') = -\frac{1}{2\pi}\partial_s \log\left|\frac{\pi}{L} \cdot \frac{x(s) - x(s')}{\sin(\pi(s-s')/L)}\right|$$

is *bounded* away from the corners, and for bounded $s, s'$. (One derivative is used to bound the difference quotient, and the second is used by the differentiation w.r.t. $s$.) Thus, the assertion of Lemma 3.6 follows. □

Thus, the only relevant term coming from $\mathbf{A}^{(3)}$ is $\mathbf{A}^{(4)}$ near a corner (and also near the diagonal). These "corner terms" are

$$\mathbf{B}_j(s,s') \equiv h(\zeta(s-s')) \cdot h(\zeta(s'-s_j))\mathbf{A}^{(3)}(s,s') .$$

Since the supports of the localizers are disjoint for different $j$, and the expressions are translation invariant, we may assume without loss of generality that $s_j = 0$ and we omit henceforth the index $j$. We now straighten the edges near the corners as follows. We let $x_\pm$ denote the two unit tangent vectors along $\Gamma$, pointing away from $s = 0$. We set $y(s) = s \cdot x_+$, when $s > 0$ and $y(s) = -s \cdot x_-$, when $s < 0$. Then we define

$$\mathbf{B}^{(0)}(s,s') = -\frac{1}{2\pi}h(\zeta(s-s')) \cdot h(\zeta(s'))\log\left|\frac{\pi}{L} \cdot \frac{y(s) - y(s')}{\sin(\pi(s-s')/L)}\right| ,$$

which is just like **B**, but with $y(s)$ replacing $x(s)$ in the quotient. We can now go to the straight coordinates, by virtue of

**Lemma 3.7.** *One has*
$$i\partial_s \left( \mathbf{B}(s,s') - \mathbf{B}^{(0)}(s,s') \right) \approx 0 \,.$$

**Proof.** The difference of the logarithms leads to a term proportional to
$$i\partial_s h(\zeta(s-s')) \cdot h(\zeta(s')) \log \left( \frac{x(s) - x(s')}{y(s) - y(s')} \right)^2 \,.$$

The chain rule of differentiation creates 2 terms, $\mathbf{T_1} + \mathbf{T_2}$, of which the first is compact, because it is localized away from the diagonal.

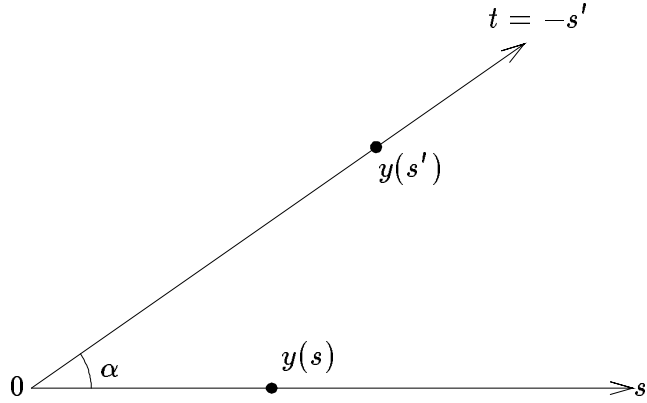

**Fig. 3**: The coordinate system near a corner of the boundary $\Gamma$.

The second term is more complicated to bound, and makes use of the geometry of a corner, cf. Fig. 3. We study first the second term when $ss' < 0$. Without loss of generality we consider only the case $s > 0, t = -s' > 0$. Denoting $\alpha$ the angle between the two tangents, we have
$$|y(s) - y(t)|^2 = s^2 + t^2 - 2st \cos(\alpha) \,. \tag{3.11}$$

**Remark.** Since we assume the corners have angles for which $|\cos(\alpha)| < 1$—*by the definition of standard domain*—it follows that $|y(s) - y(t)|^2 > d(\alpha)(s^2 + t^2)$, with $d(\alpha) > 0$.

Since $x$ is a $\mathcal{C}^2$ function, we find
$$\left| \frac{x(s) - x(t)}{y(s) - y(t)} \right|^2 = 1 + \mathcal{O}((|s| + |t|)^2)/d(\alpha) \,,$$
where the last term is $\mathcal{C}^1$. Therefore, we see that
$$i\partial_s \log \left( \frac{x(s) - x(t)}{y(s) - y(t)} \right)^2 = \mathcal{O}(1) \,. \tag{3.12}$$

Therefore, the contribution to $\mathbf{T}_2$ from the region $s \cdot t > 0$ is compact. We finally estimate the contribution from $s \cdot t < 0$ to $\mathbf{T}_2$. In this case, the two points are on the same side of the corner, and hence $|y(s) - y(t)|$ is proportional to $|s - t|$. This leads to a bound

$$\left| \frac{x(s) - x(t)}{y(s) - y(t)} \right| = 1 + \mathcal{O}(s - t) ,$$

and after differentiation, we obtain again Eq.(3.12). The proof of Lemma 3.7 is complete. □

The upshot of these calculations is that the only relevant terms modulo compact operators coming from $\mathbf{A}^{(3)}$ are

$$i\partial_s \mathbf{A}^{(3)}(s, s') \approx i\partial_s \sum_{j=1}^{N} \mathbf{B}_j^{(0)}(s, s') . \qquad (3.13)$$

Assuming again that the corner is at $s = 0$ and omitting the index $j$, we analyze

$$i\partial_s \mathbf{B}^{(0)}(s, s') \approx -\frac{i}{4\pi} h(\zeta(s - s')) \cdot h(\zeta(s')) \cdot \partial_s \log \left( \frac{\pi}{L} \cdot \frac{y(s) - y(s')}{\sin(\pi(s - s')/L)} \right)^2 ,$$

since again the term involving the derivative of $h$ is supported away from the diagonal. Finally, to simplify our task, we replace the cutoff function by a simple one, modulo compacts, and redefining $h$, if necessary. Thus, we study

$$i\partial_s \mathbf{B}^{(0)}(s, s') \approx -\frac{i}{4\pi} h(s) h(s') \cdot \partial_s \log \left( \frac{\pi}{L} \cdot \frac{y(s) - y(s')}{\sin(\pi(s - s')/L)} \right)^2 .$$

As a last step, we replace the sinus by a linear function, and thus, we study

$$i\partial_s \mathbf{B}^{(0)}(s, s') \approx -\frac{i}{4\pi} h(s) \cdot h(s') \cdot \partial_s \log \left( \frac{y(s) - y(s')}{s - s'} \right)^2 . \qquad (3.14)$$

Note now that if $s$ and $s'$ have the same sign, then the argument of the logarithm is a constant. Therefore, it suffices to consider the operator $\mathbf{B}^{(0)}$ restricted to $ss' < 0$. A straightforward calculation using Eq.(3.11) shows that in this case

$$-\frac{i}{2\pi} \partial_s \log \left| \frac{y(s) - y(s')}{s - s'} \right| = -\frac{1}{2\pi i} \left( \frac{1}{s - s'} - \frac{1}{2} \frac{1}{s + s' e^{i\alpha}} - \frac{1}{2} \frac{1}{s + s' e^{-i\alpha}} \right) \equiv \mathbf{C}_\alpha(s, s') . \qquad (3.15)$$

Thus, the reductions done so far show that

$$|i\partial_s| \mathbf{A} \approx \tfrac{1}{2} \mathbf{1} + \sum_{j=1}^{N} \mathrm{sign}(i\partial_s) h(s - s_j) h(s' - s_j) \mathbf{C}_{\alpha_j}(s - s_j, s' - s_j) , \qquad (3.16)$$

where $\alpha_j$ is the angle at the $j^{\text{th}}$ corner. We continue by analyzing the operator $\mathrm{sign}(i\partial_s) h(s) h(s') \mathbf{C}_{\alpha_j}$. Before we can do so, we want to simplify the operator $\mathrm{sign}(i\partial_s)$, defined in Eq.(3.8). This is achieved by

**Lemma 3.8.** *The operator with kernel*

$$\mathbf{K}(s,s') \;=\; \frac{i}{L}\cot\bigl(\pi(s-s')/L\bigr)h(s') - \frac{i}{\pi}h(s)\frac{1}{s-s'} \tag{3.17}$$

*is compact from $L^2(\mathbf{R})$ to $L^2(I_L)$.*

**Proof.** Recall that $s \in I_L = [-L/2, L/2]$. Then, we can write

$$\frac{i}{L}\cot\bigl(\pi(s-s')/L\bigr)h(s') \;=\; \frac{i}{\pi}\frac{1}{s-s'}h(s') + h(s')\cdot\mathcal{O}(s-s')\,.$$

The second term is clearly the kernel of a compact operator from $L^2(\mathbf{R})$ to $L^2(I_L)$. Therefore,

$$\mathbf{K}(s,s') \;\approx\; -\frac{i}{\pi}\cdot\frac{h(s)-h(s')}{s-s'} \;\equiv\; \mathbf{K}_1(s,s')\,. \tag{3.18}$$

We bound the r.h.s. of (3.18) by considering three regions:
1. The region where $s, s' \in \operatorname{supp} h$: There, $\mathbf{K}_1$ is bounded, since the $h$ are $\mathcal{C}^\infty$.
2. The region where $s' \in \operatorname{supp} h$, $s \notin \operatorname{supp} h$: Again, $\mathbf{K}_1$ is bounded.
3. The region where $s \in \operatorname{supp} h$, $s' \notin \operatorname{supp} h$: This is a non-compact piece, because $s'$ varies in $\mathbf{R}$. But then $|\mathbf{K}_1(s,s')| \leq \mathcal{O}(1/s')$, and hence the kernel is in $L^2$.

Thus, $\int_{I_L} ds \int_{-\infty}^\infty ds' |\mathbf{K}(s,s')|^2 < \infty$, and $\mathbf{K}$ is Hilbert-Schmidt. The proof of Lemma 3.8 is complete. $\square$

Using Eq.(3.8) and this last lemma, we see that near any corner,

$$\begin{aligned}
|i\partial_s|\mathbf{B}^{(0)} &= \operatorname{sign}(i\partial_s)\cdot i\partial_s\mathbf{B}^{(0)} \approx \operatorname{sign}(i\partial_s)h\mathbf{C}_\alpha h \\
&\approx h\mathbf{P}\mathbf{C}_\alpha(s,s')h + (\mathbf{K}\mathbf{C}_\alpha)h\,.
\end{aligned} \tag{3.19}$$

Here $h$, denotes the operator of multiplication by $h$, and $\mathbf{P}$ is $\operatorname{sign}(i\partial_s)$, but viewed on $L^2(\mathbf{R})$, i.e., the operator whose integral kernel is

$$\mathbf{P}(s,s') \;=\; \frac{i}{\pi}\frac{1}{s-s'}\,.$$

We shall show below that $\mathbf{C}_\alpha$ is a bounded operator on $L^2(\mathbf{R})$ and therefore (3.19) implies that

$$|i\partial_s|\mathbf{B}^{(0)} \approx h\mathbf{P}\mathbf{C}_\alpha h^*\,, \tag{3.20}$$

where $h^*$ is the multiplication by $h$ (viewed as a map from $L^2(I_L)$ to $L^2(\mathbf{R})$) and $h$ maps $L^2(\mathbf{R}) \to L^2(I_L)$.

To study $\mathbf{C}_\alpha$ on $L^2(\mathbf{R})$ it is advantageous to identify $L^2(\mathbf{R})$ with $L^2(\mathbf{R}^+) \oplus L^2(\mathbf{R}^+)$, using the map $u(s) \mapsto (u_+(s), u_-(s))$ with

$$u(s) \;=\; \begin{cases} u_+(s), & \text{when } s > 0, \\ u_-(-s), & \text{when } s < 0. \end{cases}$$

Having gone to unbounded coordinates, we can now use them for an explicit calculation. We define the self-adjoint generator, $\mathbf{D}$, of the dilatations on $\mathbf{R}^+$,

$$\left(e^{i\mathbf{D}t}f\right)(s) \;=\; e^{t/2} f(e^t s) \;.$$

This operator is diagonalized by the Mellin transformation $\mathcal{M}$, defined by

$$(\mathcal{M}f)(\lambda) \;=\; \frac{1}{\sqrt{\pi}} \int_0^\infty ds\, s^{i\lambda - 1/2} f(s) \;.$$

Note that $\mathcal{M} : L^2(\mathbf{R}^+) \to L^2(\mathbf{R})$ is unitary and $\mathcal{M}e^{i\mathbf{D}t} = e^{i\lambda t}\mathcal{M}$. With the above notation, we see that $\mathcal{M}\mathbf{C}_\alpha u$ is given by

$$(\mathcal{M}\mathbf{C}_\alpha)_\sigma(\lambda)$$
$$= -\frac{\sigma}{\sqrt{\pi}} \int_0^\infty ds\, s^{i\lambda - 1/2} \int_0^\infty \frac{ds'}{2\pi i} \Big(\frac{1}{s+s'} - \frac{1}{2}\frac{1}{s - s'e^{i\alpha}} - \frac{1}{2}\frac{1}{s - s'e^{-i\alpha}}\Big) u_{-\sigma}(s') \;,$$

where $\sigma \in \{+, -\}$. Replacing the integration variable $s$ by $ss'$ and noting that the integrand is homogeneous of degree $i\lambda + 1/2$ in $s'$, we get

$$(\mathcal{M}\mathbf{C}_\alpha u)_\sigma(\lambda) \;=\; -\sigma \int_0^\infty \frac{ds}{2\pi i} s^{i\lambda - 1/2} \Big(\frac{1}{s+1} - \frac{1}{2}\frac{1}{s - e^{i\alpha}} - \frac{1}{2}\frac{1}{s - e^{-i\alpha}}\Big)(\mathcal{M}u_{-\sigma})(\lambda) \;,$$
$$\equiv -\sigma c_\alpha(\lambda)(\mathcal{M}u_{-\sigma})(\lambda) \;.$$

Thus, $\mathbf{C}_\alpha$ becomes matrix multiplication under the Mellin transform. We next evaluate the integral $c_\alpha(\lambda)$. Note that the integrand is $\mathcal{O}(s^{-3/2})$ at infinity and $\mathcal{O}(s^{-1/2})$ near 0. Therefore, for large $R$, we find

$$c_\alpha(\lambda) \;=\; \int_{R^{-1}}^{R} \frac{ds}{2\pi i} s^{i\lambda - 1/2} \Big(\frac{1}{s+1} - \frac{1}{2}\frac{1}{s - e^{i\alpha}} - \frac{1}{2}\frac{1}{s - e^{-i\alpha}}\Big) + \mathcal{O}(R^{-1/2}) \;. \qquad (3.21)$$

The integrand is meromorphic in the annular sector $\{s \,:\, 1/R < |s| < R,\, \arg(s) \in (0, 2\pi)\}$. To evaluate the integral, we consider the contour given in Fig. 4.
The integral over the circles which are concentric around the origin contributes $\mathcal{O}(R^{-1/2})$ and the integral over the segment $1/R \le s \le R$, $\arg(s) = 2\pi$ equals $(e^{2\pi i})^{i\lambda - 1/2} c_\alpha(\lambda)$. Letting $R \to \infty$, we obtain

$$0 = -c_\alpha(\lambda) + (e^{2\pi i})^{i\lambda - 1/2} c_\alpha(\lambda) + \sum_{z=\{e^{i\alpha}, e^{i\pi}, e^{i(2\pi - \alpha)}\}} \operatorname*{Res}_{s=z}\Big(\frac{s^{i\lambda - 1/2}}{s+1} - \frac{1}{2}\frac{s^{i\lambda - 1/2}}{s - e^{i\alpha}} - \frac{1}{2}\frac{s^{i\lambda - 1/2}}{s - e^{-i\alpha}}\Big) \;.$$

This leads to

$$c_\alpha(\lambda) \;=\; -\frac{i + \sinh\big((\pi - \alpha)\lambda - i\alpha/2\big)}{2\cosh(\pi\lambda)} \;.$$

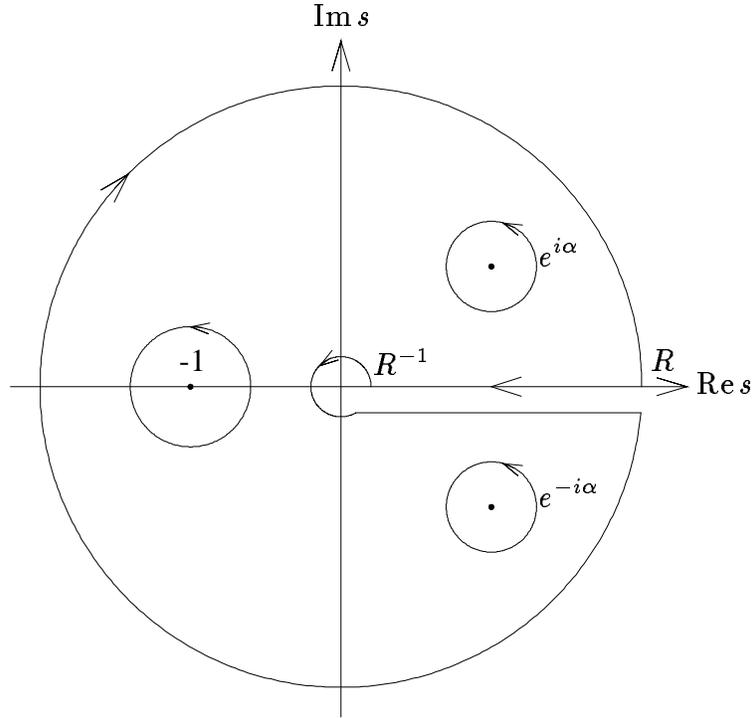

**Fig. 4**: The contour used in evaluating the integral $c_\alpha(\lambda)$ of Eq.(3.21).

Note that $c_\alpha(\lambda) = 0$ when $\alpha = \pi$. We next compute the operator **P** in the Mellin representation [Gr, D]. We find

$$(\mathcal{M}\mathbf{P}u)_\sigma(\lambda) = \sum_{\sigma'} M_{\sigma,\sigma'} (\mathcal{M}u)_{\sigma'}(\lambda),$$

where the matrix $M$ has elements

$$M_{--} = -M_{++} = -\tanh(\pi\lambda),$$
$$M_{-+} = -M_{+-} = \frac{i}{\cosh(\pi\lambda)}.$$

Altogether, we find

$$\mathcal{M}\mathbf{P}\mathbf{C}_\alpha \mathcal{M}^* = \begin{pmatrix} \dfrac{-1 + i\sinh\big((\pi-\alpha)\lambda - i\alpha/2\big)}{2\cosh^2(\pi\lambda)} & \tanh(\pi\lambda)\dfrac{i + \sinh\big((\pi-\alpha)\lambda - i\alpha/2\big)}{2\cosh(\pi\lambda)} \\ \tanh(\pi\lambda)\dfrac{i + \sinh\big((\pi-\alpha)\lambda - i\alpha/2\big)}{2\cosh(\pi\lambda)} & \dfrac{-1 + i\sinh\big((\pi-\alpha)\lambda - i\alpha/2\big)}{2\cosh^2(\pi\lambda)} \end{pmatrix}.$$

The eigenvalues $b(\alpha, \lambda)$ of $\mathcal{M}\mathbf{P}\mathbf{C}_\alpha\mathcal{M}^*$ are then

$$b_\pm(\alpha, \lambda) = \frac{-1 + i\sinh\big((\pi - \alpha)\lambda - i\alpha/2\big)}{2\cosh^2(\pi\lambda)} \pm \tanh(\pi\lambda)\frac{i + \sinh\big((\pi - \alpha)\lambda - i\alpha/2\big)}{2\cosh(\pi\lambda)} \, . \tag{3.22}$$

Using the definition of $b_\pm(\alpha, \lambda)$, we see that $b_\pm(\alpha, \lambda) = \bar{b}_\pm(\alpha, -\lambda)$. Furthermore,

$$\begin{aligned}|b_\pm(\alpha, \lambda)| &= \frac{\sqrt{1 + \sinh^2(\pi\lambda)}}{2\cosh^2(\pi\lambda)}\,\big|\sinh\big((\pi - \alpha)\lambda - i\alpha/2\big)\big|^2 \\ &= \frac{\cosh\big((\pi - \alpha)\lambda\big) - \cos\big((\pi - \alpha)/2\big)}{2\cosh(\pi\lambda)} < \frac{\cosh\big((\pi - \alpha)\lambda\big)}{2\cosh(\pi\lambda)} < \tfrac{1}{2}\,.\end{aligned} \tag{3.23}$$

We can now complete the proof of Theorem 3.1. Consider first Eqs.(3.10), (3.16), (3.19), and finally (3.20). By Eq.(3.23), we see that every corner contributes a bounded piece to $|i\partial_s|\mathbf{A}$. Combining all these estimates, we see that $|i\partial_s|\mathbf{A}$ is bounded on $\mathcal{H}_\Gamma$, which means that $\mathbf{A}$ is bounded from $\mathcal{H}_\Gamma$ to $\mathcal{H}^1_\Gamma$.

We next determine a bound on the essential spectrum of $|i\partial_s|\mathbf{A}$. In Fig. 5 we show the essential spectrum of $\tfrac{1}{2}\mathbf{1} + \mathbf{B}^{(0)}$, for one corner, i.e., the set $1/2 + b_\pm(\alpha, \lambda)$, $\lambda \in \mathbf{R}$. The curve of Fig. 5 encloses the essential spectrum of $|i\partial_s|\mathbf{A}$.

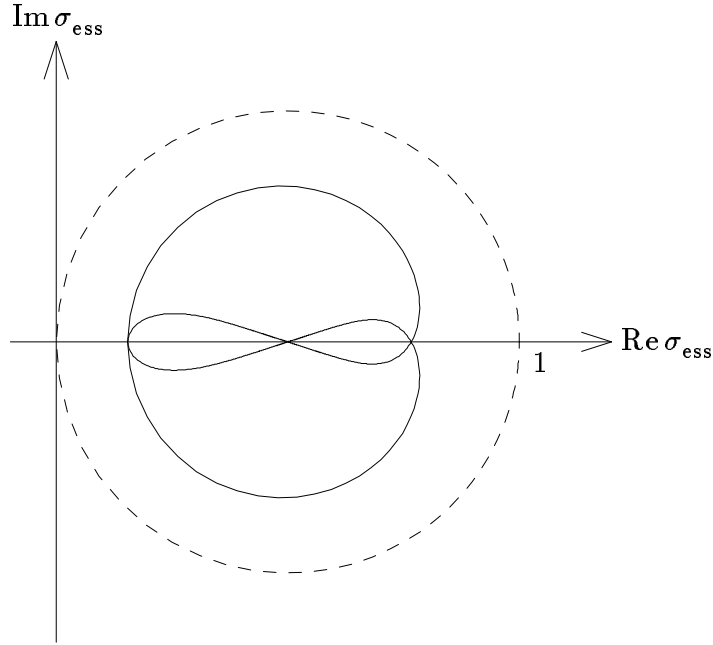

**Fig. 5**: The essential spectrum of the operator $\tfrac{1}{2}\mathbf{1} + \mathbf{B}^{(0)}$, for the case of one corner with $\alpha = 0.9 \cdot 2\pi$. Note that it lies strictly in the right half plane. In fact we show that for $\alpha$ with $|\cos(\alpha)| < 1$, it lies strictly inside the dashed circle.

Note that $\mathcal{M}\mathbf{PC}_\alpha\mathcal{M}^*$ is normal so we have the estimate

$$|(f, \mathbf{PC}_\alpha f)| \leq \rho(\alpha)\|f\|^2_{L^2(\mathbf{R})},$$

with $\rho(\alpha) = \max_{\lambda \in \mathbf{R}} |b_\pm(\alpha, \lambda)| < \frac{1}{2}$. Therefore, denoting by $hf$ the cutoff of $f$ near a corner, we find

$$|(f, h\mathbf{PC}_\alpha hf)| \leq \rho(\alpha)\|hf\|^2_{L^2(\mathbf{R})} \leq \rho(\alpha)\|f\|^2_{L^2(I_L)}.$$

Since the supports of the distinct $\mathbf{B}_j^{(0)}$ are disjoint, we have

$$|(f, (|i\partial_s|\sum_j \mathbf{B}_j^{(0)} - \mathcal{K})f)| \leq |\sum_j (f, h_j \mathbf{PC}_{\alpha_j} h_j f)|$$
$$\leq \sum_j \rho(\alpha_j)\|h_j f\|^2 \leq \max_j \rho(\alpha_j)\|f\|^2_{L^2(I_L)},$$

where $\mathcal{K}$ is the compact error term. By Weyl's theorem [K], it follows that

$$\begin{aligned}\sigma_{\text{ess}}(|i\partial_s|\mathbf{A}^{(3)}) &= \sigma_{\text{ess}}(|i\partial_s|\mathbf{A}^{(4)}) \\ &= \sigma_{\text{ess}}(|i\partial_s|\sum \mathbf{B}_j^{(0)}) \subset \{z \,:\, |z| < \max_j \rho(\alpha_j) < \tfrac{1}{2}\}.\end{aligned} \quad (3.24)$$

This implies that $0 \notin \sigma_{\text{ess}}(|i\partial_s|\mathbf{A})$.

We can now prove the Fredholm property. By the decomposition Eq.(3.10) and by Eqs.(3.13), (3.24) we have

$$|i\partial_s|\mathbf{A} = \tfrac{1}{2}\mathbf{1} + \mathbf{B}^{(0)} + \mathcal{K}, \qquad (3.25)$$

where $\|\mathbf{B}^{(0)}\| \equiv r < 1/2$, and $\mathcal{K}$ is compact. Since $\Lambda = (1 + (i\partial_s)^2)^{1/2}$ and $|i\partial_s|$ have the same asymptotic distribution of eigenvalues, their difference is compact, and therefore Eq.(3.25) implies

$$\Lambda \mathbf{A} = \tfrac{1}{2}\mathbf{1} + \mathbf{B}^{(0)} + \mathcal{K}_1. \qquad (3.26)$$

Here, and in the sequel, $\mathcal{K}_1, \ldots$ denote compact operators. The Eq.(3.26) clearly implies that $\Lambda \mathbf{A}$ is Fredholm on $\mathcal{H}_\Gamma$.

To see that its index is 0, we note that $\Lambda \mathbf{A}$ is a compact perturbation of $\tfrac{1}{2}\mathbf{1} + \mathbf{B}^{(0)}$ and hence has the same index. But the latter operator has index 0 because its kernel and that of its adjoint are trivial. The proof of Theorem 3.1 for $\beta = 0$ is complete. $\square$

Note that because the logarithmic term is real, it can only contribute to the real part of $\mathbf{A}$ and therefore the proof of Theorem 3.1 really implies

$$\begin{aligned}\Lambda \mathbf{Y} &= \tfrac{1}{2}\mathbf{1} + \mathbf{B}^{(0)} + \mathcal{K}_2, \\ \Lambda \mathbf{J} &= \mathcal{K}_3.\end{aligned} \qquad (3.27)$$

We now extend these results to $\beta \in [0, 1]$. We need some machinery to compare $\Lambda \mathbf{A}$ with $\Lambda^{1-\beta}\mathbf{A}\Lambda^\beta$.

**Definition.** Let $Z$ be a bounded operator. Then we define its Fredholm radius by

$$\rho_{\rm F}(Z) \;=\; \inf_{\mathcal{K}\,:\,\mathcal{K}\text{ compact}} \|Z + \mathcal{K}\| \;.$$

**Lemma 3.9.** *Let $Z$ be a bounded, selfadjoint operator and let $\Lambda \geq 1$ be a positive, selfadjoint, possibly unbounded operator. Assume that $Z$ maps into the domain of $\Lambda$, i.e.,*

$$\mathrm{Ran}(Z) \subset D(\Lambda) \;.$$

*Then, for all $\beta \in [0,1]$, the operator $\Lambda^{1-\beta} Z \Lambda^\beta$ is bounded, and*

$$\rho_{\rm F}(\Lambda^{1-\beta} Z \Lambda^\beta) \;\leq\; \rho_{\rm F}(\Lambda Z) \;. \tag{3.28}$$

Postponing the proof of this lemma, we continue the proof of Theorem 3.1. Consider $Z = \Lambda^{-1}\mathbf{B}^{(0)}$. Then $\rho_{\rm F}(\Lambda Z) = r < \tfrac{1}{2}$, by Eq.(3.24). The lemma implies $\rho_{\rm F}(\Lambda^{-\beta}\mathbf{B}^{(0)}\Lambda^\beta) = \rho_{\rm F}(\Lambda^{1-\beta} Z \Lambda^\beta) =\leq r < \tfrac{1}{2}$. Similarly, choosing $Z = \Lambda^{-1}\mathcal{K}_{2,3}$ we obtain $\rho_{\rm F}(\Lambda^{-\beta}\mathcal{K}_{2,3}\Lambda^\beta) = 0$. Acting with $\Lambda^{-\beta} \cdot \Lambda^\beta$ on Eq.(3.27), we get immediately

$$\begin{aligned}
\Lambda^{1-\beta}\mathbf{Y}\Lambda^\beta &= \tfrac{1}{2}\mathbf{1} + \mathbf{B}_\beta + \mathcal{K}_4 \;, \\
\Lambda^{1-\beta}\mathbf{J}\Lambda^\beta &= \mathcal{K}_5 \;,
\end{aligned} \tag{3.29}$$

where $\mathbf{B}_\beta = \Lambda^{-\beta}\mathbf{B}^{(0)}\Lambda^\beta$ has norm bounded by $r < 1/2$, and $\mathcal{K}_4, \mathcal{K}_5$ are compact, and analytic in $k$. Thus, $\Lambda^{1-\beta}\mathbf{A}\Lambda^\beta$ has the same properties as those shown for $\beta = 0$. The proof of Theorem 3.1 is complete. □

**Proof of Lemma 3.9.** The proof is an application of analytic interpolation methods. Let $F = \Lambda Z$. This is a bounded operator, and thus $Z\Lambda$ extends to the bounded operator $F^*$, which we write again as $F^* = Z\Lambda$, by a slight abuse of notation. By interpolation, the operator

$$F(\beta) \;=\; \Lambda^{1-\beta} Z \Lambda^\beta \;, \quad \beta \in [0,1] \;,$$

also extends to a bounded operator, and $\|F(\beta)\| \leq \|F\|$. Since $\Lambda^{it}$ is unitary, for real $t$, we have the same bound for $F(w)$, where $w \in S \equiv \{0 \leq \mathrm{Re}\,(w) \leq 1\}$. For $u, v \in D(\Lambda)$, the matrix elements

$$(u, F(w)v) \;=\; (\Lambda^{1-\bar{w}} u, Z\Lambda^w v)$$

are analytic in $w$ in the interior of the strip $S$ and by density, this is also true for arbitrary $u, v$. Hence, $F(w)$ is weakly analytic, and therefore norm-analytic in the interior of $S$.

Consider next the resolvent

$$G(w, z) \;=\; (z - F(w))^{-1} \;.$$

For $|z| > \|F\|$, this is an analytic function of $z$ which satisfies the identities

$$G(w,z)\Lambda^{-w} = \Lambda^{-w} G(0,z) = \Lambda^{-w}(z-F)^{-1},$$
$$\Lambda^{-(1-w)} G(w,z) = G(1,z)\Lambda^{-(1-w)} = (z-F^*)^{-1}.$$

Thus, the arguments above allow us to conclude that, $G(w,z)$ is analytic in

$$W_0 \equiv \{\operatorname{Re} w \in (0,1)\} \times \{|z| > \|F\|\}.$$

Furthermore, the matrix elements $f(w,z) = (u, G(w,z)v)$ are continuous in

$$W \equiv \{\operatorname{Re} w \in [0,1]\} \times \{|z| > \|F\|\}.$$

Choose now a $\rho > \rho_F(F)$. Then the functions $f(it, z)$ and $f(1+it, z)$ are meromorphic in $\{|z| > \rho\}$, with poles of order $\nu_j$ at points $z_j$, $j=1,\ldots,N$. Define next

$$g(w,z) = \prod_{j=1}^{N}(1 - z_j/z) \cdot e^{(w-1/2)^2} f(w,z).$$

Then $g$ is analytic in $W_0$, continuous in $W$, and analytic in $\{|z| > \rho\}$ when $w = it$ or $w = 1+it$. Furthermore, as $w \to \infty$ inside the strip $S$, we have the bound $g(w,z) = \mathcal{O}(e^{-|\operatorname{Im} w|^2})$. Therefore, the Cauchy integral yields

$$g(w,z) = \frac{1}{2\pi} \int_{-\infty}^{\infty} dt \left( \frac{g(1+it,z)}{1-w+it} + \frac{g(it,z)}{w-it} \right).$$

Thus, this analytic completion argument shows that $g$ is analytic in the envelope $S \times \{|z| > \rho\}$. Thus, $f$ is meromorphic in the same domain and thus $\sigma_{\text{ess}}(F(w)) \cap \{|z| > \rho\} = \emptyset$. Since $\rho > \rho_F$ was arbitrary, the assertion of Lemma 3.9 is proved. □

**Proof of Corollary 3.2.** By Eq.(3.2), we know that

$$\Lambda^{1/2} \mathbf{Y} \Lambda^{1/2} = \tfrac{1}{2}\mathbf{1} + \mathbf{B} + \mathcal{K}_1,$$

where $\tfrac{1}{2}\mathbf{1} + \mathbf{B} \geq \tfrac{1}{2} - r > 0$. Therefore, the subspace on which $(u, \Lambda^{1/2}\mathbf{Y}\Lambda^{1/2} u)$ is negative has finite dimension. The subspace on which $(v, \mathbf{Y}v) = (v, \Lambda^{-1/2}\Lambda^{1/2}\mathbf{Y}\Lambda^{1/2}\Lambda^{-1/2}v)$ is negative has the same dimension, and hence the Corollary 3.2 follows from the minimax principle.

**Proof of Corollary 3.3.** It clearly suffices to show that $\ker(\mathbf{A}|_{\mathcal{H}_\Gamma^{-1}}) \subset \ker(\mathbf{A}|_{\mathcal{H}_\Gamma})$, since $\mathcal{H}_\Gamma^{-1} \supset \mathcal{H}_\Gamma$. Since $\mathbf{A}^* = \bar{\mathbf{A}}$ by Eq.(2.12), we have

$$(\mathbf{A}|_{\mathcal{H}_\Gamma})^* = \bar{\mathbf{A}}|_{\mathcal{H}_\Gamma^{-1}}.$$

By Theorem 3.1, we also have $\operatorname{index}(\mathbf{A}|_{\mathcal{H}_\Gamma}) = 0$, and since it is preserved by conjugation, $\operatorname{index}(\bar{\mathbf{A}}|_{\mathcal{H}_\Gamma}) = 0$. From $\ker(\bar{\mathbf{A}}) = \overline{\ker(\mathbf{A})}$, we have

$$\dim \ker(\mathbf{A}|_{\mathcal{H}_\Gamma^{-1}}) = \dim \ker((\bar{\mathbf{A}}|_{\mathcal{H}_\Gamma})^*) = \dim \ker(\bar{\mathbf{A}}|_{\mathcal{H}_\Gamma})$$
$$= \dim \overline{\ker}(\mathbf{A}|_{\mathcal{H}_\Gamma}) = \dim \ker(\mathbf{A}|_{\mathcal{H}_\Gamma}).$$

The proof is complete. □

## 4. The relation between $\mathbf{A}_k$ and the Dirichlet boundary value problems

In this section, we establish the relations between the boundary restriction $\mathbf{A}_k$, the spectrum of $\Delta_\Omega$, and the on-shell S-matrix.

**Definition and properties of the restriction to the energy shell.** We recall that the energy shell is $F_k = \{p \in \mathbf{R}^2 \mid p^2 = k^2\}$. We define the restriction $\Sigma_k$ to the energy shell $F_k$:

$$(\Sigma_k \psi)(p) = \int_{\mathbf{R}^2} d^2 y\, e^{-ipy} \psi(y)\,, \quad \text{when } p \in F_k\,,$$

$$(\Sigma_k^* \chi)(x) = \int_{F_k} d\mu(p) e^{ipx} \chi(p)\,, \quad \text{when } x \in \mathbf{R}^2\,.$$

Here, $d\mu = (4\pi)^{-1} d\varphi$, where $\varphi$ is the angle on the circle $F_k$. We will use the following facts about these operators, which follow easily from the definition:

$$\begin{aligned}\Sigma_k^* &: L^2(F_k) \to H^\beta_{\text{loc}}(\mathbf{R}^2)\,, \text{ for all } \beta \geq 0\,,\\ \Sigma_k &: H^{-\beta}_{\text{comp}}(\mathbf{R}^2) \to L^2(F_k)\,, \text{ for all } \beta \geq 0\,.\end{aligned} \quad (4.1)$$

Furthermore, $\Sigma_k^*$ has trivial kernel,

$$\ker(\Sigma_k^*) = \{0\}\,. \quad (4.2)$$

We can now combine the actions of $\gamma$ (defined in Sect.2.) and $\Sigma$ into the operator $\mathcal{L}$:

**Definition and properties of $\mathcal{L}_k$ and $\mathcal{L}_k^*$.** We define

$$\mathcal{L}_k = \Sigma_k \gamma^*\,.$$

The properties of $\gamma$ and $\Sigma_k$ then imply

$$\begin{aligned}\mathcal{L}_k &: \mathcal{H}_\Gamma \to L^2(F_k)\,,\\ \mathcal{L}_k^* &: L^2(F_k) \to \mathcal{H}_\Gamma\,.\end{aligned} \quad (4.3)$$

Since $F_k$ is bounded, it follows from the definitions that one has the stronger properties

$$\begin{aligned}\mathcal{L}_k^* &: L^2(F_k) \to \mathcal{H}_\Gamma^1\,,\\ \mathcal{L}_k &: \mathcal{H}_\Gamma^{-1} \to L^2(F_k)\,.\end{aligned} \quad (4.4)$$

The next lemma relates $\mathcal{L}_k$ to the imaginary part $\mathbf{J}_k$ of $\mathbf{A}_k$.

**Lemma 4.1.** *Let $\Omega$ be a standard domain. For all $k > 0$ one has the identity*

$$\mathbf{J}_k = \pi \mathcal{L}_k^* \mathcal{L}_k\,. \quad (4.5)$$

**Proof.** Let $u \in \mathcal{H}_\Gamma$. By definition, $\mathbf{J} = \operatorname{Im}(\mathbf{A}) = (2i)^{-1}(\mathbf{A} - \mathbf{A}^*)$. Therefore,

$$(u, \mathbf{J}u) = \operatorname{Im} \int_\Gamma d\sigma(z)\, d\sigma(z')\, \bar{u}(z)\, G_k^+(z - z')\, u(z')\,.$$

Since $\operatorname{Im}\left(\left(-\Delta - (k^2 + i0)\right)^{-1}\right) = \pi\,\delta(-\Delta - k^2) \equiv \mathbf{C}$, this implies

$$(u, \mathbf{J}u) = \int_\Gamma d\sigma(z)\, d\sigma(z')\, \bar{u}(z)\, \mathbf{C}(z - z')\, u(z')\,. \tag{4.6}$$

Going to Fourier transforms, we see that this implies

$$(u, \mathbf{J}u) = \int d^2 p \int_\Gamma d\sigma(z)\, d\sigma(z')\, e^{-ipz'} \bar{u}(z')\delta(p^2 - k^2) e^{ipz} u(z)\,. \tag{4.7}$$

Going back to the definitions of $\gamma$ and $\Sigma$, one sees that $(u, \mathbf{J}u) = \pi(\gamma^* u, \Sigma^* \Sigma \gamma^* u)$, so that (4.5) follows. The proof is complete. $\square$

**Lemma 4.2.** *Let $k > 0$. The following kernels coincide:*

$$\ker(\mathbf{A}_k) = \ker(\mathbf{A}_k^*) = \ker(\mathbf{J}_k|_{\mathcal{H}_\Gamma^{-1/2}}) = \ker(\mathcal{L}_k|_{\mathcal{H}_\Gamma^{-1/2}})\,. \tag{4.8}$$

**Proof.** The properties of $\ker(\mathbf{A}_k)$ are described in Corollary 3.3, and Eq.(2.12) says that $(\mathbf{A}|_{\mathcal{H}_\Gamma})^* = \bar{\mathbf{A}}|_{\mathcal{H}_\Gamma^{-1}}$. Therefore $\ker(\mathbf{A}) = \ker(\mathbf{A}^*)$.

By definition, we have $\mathbf{J} = \operatorname{Im}(\mathbf{A})$. We first show that $\mathbf{A}u = 0$ implies $\mathbf{J}u = 0$. Indeed, $\mathbf{A}u = 0$ implies $u \in \mathcal{H}_\Gamma$ and $\operatorname{Im}(u, \mathbf{A}u) = 0$, that is, $(u, \mathbf{J}u) = 0$. Since $\mathbf{J} = \pi \mathcal{L}^* \mathcal{L}$ this means $\|\mathcal{L}u\| = 0$. Thus, $\mathcal{L}u = 0$ and therefore $\mathbf{J}u = 0$, as asserted.

Assume next $\mathbf{J}u = 0$ and $u \in \mathcal{H}_\Gamma^{-1/2}$. Then, by Eq.(2.1), one has $\gamma^* u \in H^{-1}_{\mathrm{comp}}(\mathbf{R}^2)$. On the other hand, $\frac{1}{\pi}(u, \mathbf{J}u) = (u, \mathcal{L}^* \mathcal{L}u) = \|\mathcal{L}u\|^2 = 0$, and therefore $\mathcal{L}u = 0$. Denoting the Fourier transform of $\psi$ by $\hat{\psi}$, we consider

$$\widehat{(\gamma^* u)}(p) = \int_\Gamma d\sigma(z) e^{-ipz} u(z)\,.$$

Since $\widehat{(\gamma^* u)}$ is the Fourier transform of a distribution with compact support, it is entire and bounded on $\mathbf{R}^2$. Since $\mathcal{L}u = 0$, we find $\widehat{(\gamma^* u)}(p) = 0$ when $p$ is on the energy shell $F_k$. Thus, we can divide by $p^2 - k^2$ and we see that

$$\widehat{(\mathbf{G}_k^+ u)}(p) = \frac{1}{p^2 - k^2} \widehat{(\gamma^* u)}(p)$$

is defined and is in $L^2(\mathbf{R}^2)$, since $\gamma^* u \in H^{-1}_{\mathrm{comp}}(\mathbf{R}^2)$. Note now that $\mathbf{G}_k^+ u$ is a solution of the Helmholtz equation and is in $L^2(\mathbf{R}^2)$. Therefore, it must vanish at infinity and hence on *all* of $\Omega^c$. *(Here, we make use of the assumption that $\Omega^c$ is connected.)* But this means $\mathbf{A}_k u = 0$. The proof is complete. $\square$

**Remark.** It follows from the proof that $\mathbf{A}_k u = 0$ implies that

$$\psi = \mathbf{G}_k^+ u = \mathbf{G}_k^- u, \tag{4.9}$$

vanishes on the complement of $\Omega$.

We can now establish a resolvent formula for the Dirichlet problem. We use the notation $G = (-\Delta - z)^{-1}$ and $G_\Omega = (-\Delta_\Omega - z)^{-1}$.

**Theorem 4.3.** *Let $\Omega$ be a standard domain, and let $z = k^2 + i0$. Then*

$$G_\Omega \oplus G_{\Omega^c} = G - G\gamma^* \mathbf{A}_k^{-1} \gamma G. \tag{4.10}$$

**Proof.** We take $z = k^2 + i\eta, \eta > 0$. Let $\psi \in L^2(\mathbf{R}^2)$ and define $\varphi = (G - G\gamma^*(\gamma G\gamma^*)^{-1} G\gamma)\psi$. The operator $\gamma G\gamma^*$ is Fredholm from $\mathcal{H}_\Gamma^{-1/2}$ to $\mathcal{H}_\Gamma^{1/2}$, by the Remark 4 following the Corollary 3.3. Note further that

$$\operatorname{Im} \gamma G\gamma^* = \gamma \frac{\eta}{(-\Delta - k^2)^2 + \eta^2} \gamma^*,$$

so that it is strictly positive by Eq.(2.2). Therefore, $(\gamma G\gamma^*)^{-1}$ exists and maps $\mathcal{H}_\Gamma^{1/2}$ to $\mathcal{H}_\Gamma^{-1/2}$. Combining this with Eqs.(2.1) and (2.5) we see that $\varphi \in H^1(\mathbf{R}^2)$. By construction, $(-\Delta - z)\varphi = \psi$ on $\mathbf{R}^2 \setminus \Gamma$ and $\gamma\varphi = 0$. Thus, the r.h.s. of Eq.(4.10) is equal to $G_\Omega \oplus G_{\Omega^c}$. The proof is completed by noting that

$$\lim_{\eta \downarrow 0} \gamma G\gamma^* = \mathbf{A}_k.$$

$\square$

**Lemma 4.4.** *Let $\mathbf{P}$ denote the orthogonal projection onto $\ker(\mathbf{A}_{k_0}) \subset \mathcal{H}_\Gamma$. Then the operator*

$$\left. \partial_k \mathbf{P} \mathbf{A}_k \mathbf{P} \right|_{k=k_0} \tag{4.11}$$

*is positive (on $\ker(\mathbf{A}_{k_0})$). The residue of $\mathbf{A}_k^{-1}$ at $k_0 > 0$ is given by*

$$\operatorname{res} \mathbf{A}_k^{-1} = \mathbf{P} \left( \left. \partial_k \mathbf{P} \mathbf{A}_k \mathbf{P} \right|_{k=k_0} \right)^{-1} \mathbf{P}. \tag{4.12}$$

**Proof.** Let $u \in \ker(\mathbf{A}_{k_0})$, $u \not\equiv 0$. We denote $f = \gamma^* u$. We have already argued above that $\mathbf{PA}_k\mathbf{P}$ is analytic. Using scalar products in $\mathcal{H}_\Gamma$ and in $L^2(\mathbf{R}^2)$, as adequate, we have

$$\begin{aligned}
\partial_k(u, \mathbf{A}_k u)\Big|_{k=k_0} &= \lim_{k \to k_0} \left(u, \gamma \frac{G(k^2+i0) - G(k_0^2+i0)}{k-k_0} \gamma^* u\right) \\
&= \lim_{k \to k_0} \lim_{\varepsilon,\varepsilon_0 \downarrow 0} \left(u, \gamma \frac{G(k^2+i\varepsilon) - G(k_0^2+i\varepsilon_0)}{k-k_0} \gamma^* u\right) \\
&= \lim_{k \to k_0} \lim_{\varepsilon,\varepsilon_0 \downarrow 0} \Big(G(k^2-i\varepsilon)f, \\
&\qquad \frac{((-\Delta - k_0^2 - i\varepsilon_0) - (-\Delta - k^2 - i\varepsilon))}{k-k_0} G(k_0^2+i\varepsilon_0)f\Big) \\
&= \lim_{k \to k_0} \lim_{\varepsilon,\varepsilon_0 \downarrow 0} \left(G(k^2-i\varepsilon)f, \left(k+k_0 + i\frac{\varepsilon - \varepsilon_0}{k-k_0}\right)G(k_0^2+i\varepsilon_0)f\right) \equiv X .
\end{aligned}$$

By the remark after Lemma 4.2 we know that $\mathbf{G}_{k_0}^+ u$ vanishes in $\Omega^c$. Therefore, $G(k_0^2 + i\varepsilon_0)f \to \mathbf{G}_{k_0}^+ u$, weakly in $L^2(\mathbf{R}^2)$, and it follows that

$$X = \lim_{k \to k_0} \lim_{\varepsilon \downarrow 0} \left(G(k^2-i\varepsilon)f, \left(k+k_0 + i\frac{\varepsilon}{k-k_0}\right)\mathbf{G}_{k_0}^+ u\right) .$$

Noting again that $\mathbf{G}_{k_0}^+ u$ vanishes in $\Omega^c$, and furthermore that $G(k^2-i\varepsilon)f \to \mathbf{G}_k^- u$ in $L^2_{\text{loc}}(\mathbf{R}^2)$, we can get rid of the limit $\varepsilon \downarrow 0$. Thus, $X$ is equal to

$$\lim_{k \to k_0} \left(\mathbf{G}_k^- u, (k+k_0)\mathbf{G}_{k_0}^+ u\right) = 2k_0 \left(\mathbf{G}_{k_0}^- u, \mathbf{G}_{k_0}^+ u\right) . \tag{4.13}$$

Since $\mathbf{A}_{k_0} u = 0$, we know by Eq.(4.9) that $\mathbf{G}_{k_0}^- u = \mathbf{G}_{k_0}^+ u$. Therefore, (4.13) is equal to

$$2k_0(\mathbf{G}_{k_0}^+ u, \mathbf{G}_{k_0}^+ u) = 2k_0 \|\mathbf{G}_{k_0}^+ u\|^2 > 0 .$$

The last inequality follows from Eq.(2.9). The proof of the first statement of Lemma 4.4 is complete.

To prove the second part, we define $\mathbf{Q} = \mathbf{1} - \mathbf{P}$. Then,

$$\mathbf{A}_k = (\mathbf{PA}_k\mathbf{P} \oplus \mathbf{QA}_k\mathbf{Q}) + (\mathbf{PA}_k\mathbf{Q} + \mathbf{QA}_k\mathbf{P}) .$$

By Lemma 4.2, we have $\ker(\mathbf{A}_{k_0}^*) = \ker(\mathbf{A}_{k_0})$ and therefore, $\mathbf{A}_{k_0}\mathbf{P} = 0$, and $\mathbf{PA}_{k_0} = (\mathbf{A}_{k_0}^*\mathbf{P})^* = 0$. Furthermore, $\mathbf{PA}_k\mathbf{Q}$ and $\mathbf{QA}_k\mathbf{P}$ are analytic in $k$, so that

$$\|\mathbf{PA}_k\mathbf{Q} + \mathbf{QA}_k\mathbf{P}\| = \mathcal{O}(k-k_0) .$$

Letting $\varepsilon = k - k_0$, we find, in matrix notation,

$$\mathbf{A}_k = \begin{pmatrix} (k-k_0)\mathbf{PA}_{k_0}'\mathbf{P} & 0 \\ 0 & \mathbf{QA}_{k_0}\mathbf{Q} \end{pmatrix} + \begin{pmatrix} \mathcal{O}(\varepsilon^2) & \mathcal{O}(\varepsilon) \\ \mathcal{O}(\varepsilon) & \mathcal{O}(\varepsilon) \end{pmatrix} .$$

A simple calculation leads to

$$(\mathbf{A}_k)^{-1} = \begin{pmatrix} ((k-k_0)\mathbf{P}\mathbf{A}'_{k_0}\mathbf{P})^{-1} & 0 \\ 0 & 0 \end{pmatrix} + \mathcal{O}(1).$$

The proof of Lemma 4.4 is complete. $\square$

**Lemma 4.5.** *If $u \in \ker(\mathbf{A}_{k_0})$ then $\psi = \mathbf{G}_k^{\pm} u$ is an eigenfunction of $-\Delta_\Omega$ with eigenvalue $k_0^2$. This correspondence is bijective, i.e.,*

$$\dim \ker(-\Delta_\Omega - k_0^2) = \dim \ker(\mathbf{A}_{k_0}).$$

**Proof.** Let $R_{k_0}$ denote the residue of $\mathbf{A}_k^{-1}$ at $k = k_0$. By Lemma 4.4, we have $\operatorname{Ran}(R_{k_0}) = \ker(\mathbf{A}_{k_0})$. The spectral projection onto the eigenspace $-\Delta_\Omega$ corresponding to $k_0^2$ is given by the residue of the resolvent. By the resolvent formula, this is equal to

$$G_{k_0}^{\pm} \gamma^* R_{k_0} \gamma G_{k_0}^{\mp}.$$

Since $G_{k_0}^{\pm} \gamma^*$ is injective by Eq.(2.9), the assertion follows because $R_{k_0}$ is positive. $\square$

**Proposition 4.6.** *Let $k > 0$. Then the S-matrix, restricted to the energy shell $F_k$, is given by*

$$\mathbf{S}_k = 1 - 2\pi i \mathcal{L}_k \mathbf{A}_k^{-1} \mathcal{L}_k^*. \tag{4.14}$$

**Remark.** By Eq.(4.4), $\mathcal{L}^*$ maps to $\mathcal{H}_\Gamma^1$. Therefore, by Theorem 3.1 and Lemma 4.5, $\mathbf{A}^{-1}\mathcal{L}^*$ is a bounded operator when $k^2 \notin \sigma(-\Delta_\Omega)$. Furthermore, this extends to all $k > 0$ by Lemma 4.2. Thus, $\mathbf{S}_k$ is defined on $L^2(F_k)$ for all $k > 0$.

**Proof.** We apply the resolvent formula (4.10). As is well known, see e.g., [N], taking limits in Eq.(1.1), leads, for $k, k' \in \mathbf{R}^2$, to

$$\langle k | \mathbf{S} | k' \rangle = \delta(k - k') - 2\pi i \delta(k^2 - k'^2) \langle k | \mathbf{T}_{|k|} | k' \rangle, \tag{4.15}$$

where $\mathbf{T}$ is the T-matrix. It is defined as the solution of

$$G_\Omega \oplus G_{\Omega^c} = G - G\mathbf{T}_k G, \tag{4.16}$$

when $z = k^2 + i0$. By the resolvent formula, one obtains

$$\mathbf{T}_k = (\gamma G(k^2 + i0)\gamma^*)^{-1} = \mathbf{A}_k^{-1}. \tag{4.17}$$

Since the restriction $\Sigma_k$ to the energy shell satisfies $\mathcal{L}_k = \Sigma_k \gamma^*$, substitution of (4.17) into (4.15) leads to the desired result. The proof of Proposition 4.6 is complete. $\square$

**Proof of Lemma 1.1.** We use the representation Eq.(4.14) for the S-matrix. The unitarity follows from Lemma 4.1 by a simple calculation. We next show that the spectrum of $\mathbf{S}_k$ can only accumulate at 1. To see this, consider $\mathcal{L}\mathbf{A}^{-1}\mathcal{L}^*$. By the remark following Proposition 4.6, $\mathbf{A}^{-1}\mathcal{L}^*$ is bounded, and by Eq.(4.4), $\mathcal{L}$ is compact as a map from $\mathcal{H}_\Gamma$ to $L^2(F_k)$. Hence $\mathbf{S}_k - 1$ is compact. We finally show that the eigenvalues accumulate at 1 only from below. By Eq.(4.14), we have

$$\operatorname{Im} \mathbf{S}_k = -2\pi\mathcal{L}\operatorname{Re}(\mathbf{A}^{-1})\mathcal{L}^* = -2\pi\mathcal{L}(\mathbf{A}^*)^{-1}\mathbf{Y}\mathbf{A}^{-1}\mathcal{L}^* .$$

We denote by $\mathbf{Y}_+$ the positive part of $\mathbf{Y}$ and we let $\mathbf{Y}_- = \mathbf{Y} - \mathbf{Y}_+$. Then,

$$-\operatorname{Im}(\mathbf{S}_k) = 2\pi\mathcal{L}(\mathbf{A}^*)^{-1}\mathbf{Y}_+\mathbf{A}^{-1}\mathcal{L}^* + 2\pi\mathcal{L}(\mathbf{A}^*)^{-1}\mathbf{Y}_-\mathbf{A}^{-1}\mathcal{L}^* \equiv \mathbf{I}_p + \mathbf{I}_f .$$

The operator $\mathbf{I}_p$ is positive by construction and $\mathbf{I}_f$ is finite rank, by Corollary 3.2. Note that the sum of two such operators can have at most as many negative eigenvalues as the rank of the second one, as follows by writing the eigenvalues as the solutions of a minimax principle:

$$\lambda_{n+1} = \inf_{E,\dim E=n} \sup_{\psi\in E, \|\psi\|=1} (\psi, (\mathbf{I}_p + \mathbf{I}_f)\psi) .$$

Indeed, if $\ell$ is the rank of $\mathbf{I}_f$, we can find a $\psi$ orthogonal to the range of $\mathbf{I}_f$ as soon as $\dim E > \ell$, and then the supremum above is non-negative. Hence there can be at most $\ell$ negative eigenvalues, as asserted. Thus, we have shown that the number of scattering phases in the upper half plane is bounded by the rank of $\mathbf{I}_f$, and is finite. The proof of Lemma 1.1 is complete. □

## 5. Proof of the Main Theorem by a variational formula

In this section, we prove the Main Theorem and Theorem 1.3.

**Proof of Theorem 1.3.** This proof is relatively easy, because in this case the existence of the eigenfunction of $\mathbf{S}_k$ is part of the assumption. Assume $\mathbf{S}_k\chi = \chi, \chi \neq 0$, for some $\chi \in L^2(F_k)$. Let $P$ be the orthogonal projection in $\mathcal{H}_\Gamma$ onto $\ker(\mathbf{A}_k)$ and set $Q = 1 - P$. Then, by Proposition 4.6 and the remark following it, we have

$$\mathcal{L}_k Q(Q\mathbf{A}_k Q)^{-1} Q\mathcal{L}_k^*\chi = 0 . \tag{5.1}$$

We next show that $\mathcal{L}_k^*\chi = 0$. Indeed, $\ker(\mathcal{L}_k) = \ker(\mathbf{A}_k)$, by Lemma 4.2, and, by the construction of $Q$, Eq.(5.1) implies $(Q\mathbf{A}_k Q)^{-1}Q\mathcal{L}_k^*\chi = 0$. A similar reasoning then implies $Q\mathcal{L}_k^*\chi = 0$, and finally $\mathcal{L}_k^*\chi = 0$. But this means that $\gamma\Sigma_k^*\chi = 0$, by the definition of $\mathcal{L}_k$. We now claim that

$$\psi(x) = (\Sigma_k^*\chi)(x) = \int_{F_k} d\mu(p) e^{ipx}\chi(p) \tag{5.2}$$

is the desired eigenfunction. First $\psi \not\equiv 0$ because $\Sigma_k^*$ is one-to-one by Eq.(4.2). Clearly, $\psi$ solves the Helmholtz equation in all of $\mathbf{R}^2$ by construction. Since $\gamma\Sigma_k^*\chi = 0$, it also satisfies the Dirichlet boundary condition. Furthermore, it cannot vanish on an open set, because of unique continuation. Applying the Schwarz inequality to the integral in Eq.(5.2), we see that $\psi$ is uniformly bounded. This completes the proof of Theorem 1.3 for case of a simple eigenvalue. In the case of an eigenvalue with multiplicity $M$, one repeats the above calculation for $M$ linearly independent vectors $\chi_j$. Using again Eq.(4.2), the Eq.(5.2) produces $M$ independent eigenvectors. The proof Theorem 1.3 is complete. $\square$

**Remark.** It follows from this proof that $k^2 \notin \sigma(-\Delta_\Omega)$ implies $\ker(\mathcal{L}_k^*) = \{0\}$.

**Proof of the first half of the Main Theorem.** Here, we show that existence of eigenvectors implies convergence of eigenphases. We are going to use a minimax principle on the cotangents of the scattering phases. We use the definition of the $\vartheta_j$ from Eq.(1.2). Let $E_n \subset \mathcal{H}_\Gamma$ denote an $n$ dimensional subspace of $\mathcal{H}_\Gamma$.

**Theorem 5.1.** *Let $\Omega$ be a standard domain and let $k^2 \notin \sigma(-\Delta_\Omega)$. For $j \geq 0$, the cotangent of the scattering phase $\vartheta_j(k)$ of $\mathbf{S}_k$ is given by*

$$\cot \vartheta_j(k) \;=\; \inf_{E_{j+1}} \; \sup_{u \in E_{j+1}} \; \frac{(u, \mathbf{Y}_k u)}{(u, \mathbf{J}_k u)} \;. \qquad (5.3)$$

**Proof.** It is useful to consider the Cayley transform $\mathbf{X}_k$ of $\mathbf{S}_k$, given by

$$\mathbf{X}_k \;=\; i(1 + \mathbf{S}_k)(1 - \mathbf{S}_k)^{-1} \;. \qquad (5.4)$$

For $k^2 \notin \sigma(-\Delta_\Omega)$, the Theorem 1.3 says that $1 \notin \sigma(\mathbf{S}_k)$. Therefore $\mathbf{X}_k$ has dense domain $D(\mathbf{X}_k) = \mathrm{Ran}(1 - \mathbf{S}_k)$. Since $\mathbf{S}_k$ is unitary, it follows that $\mathbf{X}_k$ is self-adjoint [Ru, Theorem 13.19]. Using the spectral mapping theorem we obtain:

**Lemma 5.2.** *Let $\Omega$ be a standard domain, and let $k^2 \notin \sigma(-\Delta_\Omega)$. Let $\vartheta \in (0, \pi)$ be given. Then $e^{-2i\vartheta}$ is an eigenvalue of $\mathbf{S}_k$ of multiplicity $M$ if and only if $\cot \vartheta$ is an eigenvalue of multiplicity $M$ of $\mathbf{X}_k$.*

We introduce the polar decomposition of $\mathcal{L}$:

**Lemma 5.3.** *For $k^2 \notin \sigma(-\Delta_\Omega)$ there is a unitary operator $U_k : \mathcal{H}_\Gamma \to L^2(F_k, d\mu)$ for which*

$$\mathcal{L}_k \;=\; U_k |\mathcal{L}_k| \;, \quad \mathbf{J}_k^{1/2} \;=\; \sqrt{\pi} |\mathcal{L}_k| \;. \qquad (5.5)$$

**Proof.** The existence of a polar decomposition is well known [K, 6.2.7]. We have already shown that $k^2 \notin \sigma(-\Delta_\Omega)$ implies $\ker(\mathcal{L}_k) = \ker(\mathcal{L}_k^*) = \{0\}$, and therefore $U_k$ is not only a partial isometry but in fact unitary. The proof is complete. $\square$

We continue the proof of Theorem 5.1. The two lemmas above allow us to give another characterization of $\cot \vartheta_j(k)$, which we will use to derive Eq.(5.3). Having established that only a finite number of scattering phases are in $(\pi/2, \pi)$, we first observe that the spectral mapping theorem implies that $\mathbf{X}_k$ is bounded below, and has a finite number of negative eigenvalues.

Since $\mathbf{X}_k$ is self-adjoint and bounded below, the usual minimax principle [RS, Vol. IV Theorem 13.1] says that

$$\cot \vartheta_j(k) = \inf_{E'_{j+1} \subset D(\mathbf{X}_k)} \sup_{f \in E'_{j+1}} \frac{(f, \mathbf{X}_k f)}{(f, f)}, \tag{5.6}$$

where the infimum is taken over the $j+1$ dimensional subspaces $E'_{j+1}$ of $D(\mathbf{X}_k) \subset L^2(F_k)$ (and the supremum only over non-zero $f$). We now show, through a straightforward calculation, that Eq.(5.6) implies (5.3). If $f$ is in $D(\mathbf{X}_k)$, then, by the definition of $\mathbf{X}_k$ and the representation (4.14) of $\mathbf{S}_k$, we have

$$f = 2\pi i \mathcal{L}_k \mathbf{A}_k^{-1} \mathcal{L}_k^* \chi,$$

where $\chi \in L^2(F_k)$. Therefore,

$$\mathbf{X}_k f = 2i(1 - i\pi \mathcal{L}_k \mathbf{A}_k^{-1} \mathcal{L}_k^*) \chi. \tag{5.7}$$

We omit the index $k$ in the following calculations, and we consider only $j = 0$, to simplify the notation. Combining Eq.(5.6) with (5.7), we see that

$$\cot \vartheta_0(k) = \inf_{\chi \in L^2(F_k)} \frac{(2\pi i \mathcal{L} \mathbf{A}^{-1} \mathcal{L}^* \chi, 2i(1 - i\pi \mathcal{L} \mathbf{A}^{-1} \mathcal{L}^*) \chi)}{(2\pi i \mathcal{L} \mathbf{A}^{-1} \mathcal{L}^* \chi, 2\pi i \mathcal{L} \mathbf{A}^{-1} \mathcal{L}^* \chi)}.$$

Using the polar decomposition Eq.(5.5), $\mathcal{L} = U|\mathcal{L}| = \pi^{-1/2} U \mathbf{J}^{1/2}$, we can rewrite this as

$$\cot \vartheta_0(k) = \inf_{u \in \mathcal{H}_\Gamma} \frac{(\mathbf{J}^{1/2} \mathbf{A}^{-1} \mathbf{J}^{1/2} u, (1 - i \mathbf{J}^{1/2} \mathbf{A}^{-1} \mathbf{J}^{1/2}) u)}{(\mathbf{J}^{1/2} \mathbf{A}^{-1} \mathbf{J}^{1/2} u, \mathbf{J}^{1/2} \mathbf{A}^{-1} \mathbf{J}^{1/2} u)}.$$

We next set $v = \mathbf{A}^{-1} \mathbf{J}^{1/2} u$. Then,

$$\cot \vartheta_0(k) = \inf_{u \in \mathcal{H}_\Gamma} \frac{(\mathbf{J}^{1/2} v, u - i \mathbf{J}^{1/2} v)}{(\mathbf{J}^{1/2} v, \mathbf{J}^{1/2} v)}$$

$$= \inf_{u \in \mathcal{H}_\Gamma} \frac{(v, \mathbf{J}^{1/2} u - i \mathbf{J} v)}{(v, \mathbf{J} v)}$$

$$= \inf_{u \in \mathcal{H}_\Gamma} \frac{(v, \mathbf{A} v - i \mathbf{J} v)}{(v, \mathbf{J} v)} = \inf_{u \in \mathcal{H}_\Gamma} \frac{(v, \mathbf{Y} v)}{(v, \mathbf{J} v)}.$$

Since $\mathbf{A}_k$ is bounded, we have $\ker\big((\mathbf{A}_k^{-1})^*\big) = \{0\}$. For $k^2 \notin \sigma(-\Delta_\Omega)$, Lemma 4.5 and Lemma 4.2 imply $\ker(\mathbf{J}_k) = \{0\}$, and we find that $\ker\big(\mathbf{J}^{1/2}(\mathbf{A}^{-1})^*\big) = \{0\}$. Therefore, $\mathbf{A}^{-1} \mathbf{J}^{1/2} \mathcal{H}_\Gamma$ is dense in $\mathcal{H}_\Gamma$ and

$$\inf_{u \in \mathcal{H}_\Gamma} \frac{(v, \mathbf{Y} v)}{(v, \mathbf{J} v)} = \inf_{v \in \mathcal{H}_\Gamma} \frac{(v, \mathbf{Y} v)}{(v, \mathbf{J} v)}.$$

The proof of Theorem 5.1 is complete. □

We continue the proof of the first half of the Main Theorem. We consider a $k_0 > 0$, for which $k_0^2 \in \sigma(-\Delta_\Omega)$ is an $M$-fold eigenvalue and we denote by $\mathbf{P}$ the orthogonal projection onto $\ker(\mathbf{A}_{k_0}) \subset \mathcal{H}_\Gamma$. This kernel is $M$-dimensional by Lemma 4.5.

Let next $u \in \ker(\mathbf{A}_{k_0})$. We will show that there is a $C > 0$ for which

$$\frac{(u, \mathbf{Y}_k u)}{(u, \mathbf{J}_k u)} \leq -\frac{C}{k_0 - k}, \tag{5.8}$$

when $k < k_0$. Letting $k \uparrow k_0$ and observing that (5.8) holds for every $u \in \ker(\mathbf{A}_{k_0})$, the first half of the Main Theorem follows. In order to show Eq.(5.8), we note that since $\mathbf{P}\mathbf{A}_k\mathbf{P}$ is in fact analytic in $k$, because the integral kernel of $\mathbf{A}_k$ is analytic and $\mathbf{P}$ is finite dimensional by Theorem 3.1, we have

$$(u, \mathbf{A}_k u) = (u, \mathbf{P}\mathbf{A}_k\mathbf{P}u) = \left(u, \mathbf{P}\left((k - k_0)\mathbf{A}'_{k_0} + \mathcal{O}((k - k_0)^2)\right)\mathbf{P}u\right)$$
$$= (k - k_0)(u, \mathbf{P}\mathbf{A}'_{k_0}\mathbf{P}u) + \mathcal{O}((k - k_0)^2)\|u\|^2.$$

By Lemma 4.4, and since the kernel of $\mathbf{A}_{k_0}$ is finite dimensional, there is a $C_1 > 0$ for which

$$(u, \mathbf{P}\mathbf{A}'_{k_0}\mathbf{P}u) \geq C_1 \|u\|^2.$$

Therefore, when $k < k_0$, we have

$$(u, \mathbf{Y}_k u) = \operatorname{Re}(u, \mathbf{A}_k u) = (k - k_0)(u, \mathbf{P}\mathbf{A}'_{k_0}\mathbf{P}u) + \mathcal{O}((k - k_0)^2)\|u\|^2$$
$$\leq C_1(k - k_0)\|u\|^2 + \mathcal{O}((k - k_0)^2)\|u\|^2,$$
$$(u, \mathbf{J}_k u) = \operatorname{Im}(u, \mathbf{A}_k u) \leq C_2(k - k_0)^2.$$

Therefore, using $(u, \mathbf{J}_k u) \geq 0$, and $k < k_0$, we see that the quotient satisfies

$$\frac{(u, \mathbf{Y}_k u)}{(u, \mathbf{J}_k u)} \leq \frac{C_1(k - k_0)\|u\|^2 \cdot (1 + \mathcal{O}(k - k_0))}{(u, \mathbf{J}_k u)} \leq \frac{C_1}{C_2} \cdot \frac{1}{k - k_0}(1 + \mathcal{O}(k - k_0)),$$

from which the assertion Eq.(5.8) follows at once.

**The second half of the proof of the Main Theorem.** Here, we assume that, as $k \uparrow k_0$, there are exactly $M$ eigenvalues $e^{-2i\vartheta_j(k)}$ of $\mathbf{S}_k$ which converge to 1 from the upper half plane and show that $k_0^2$ is an $M$-fold eigenvalue of $-\Delta_\Omega$.

By the variational principle, there is, for each $k$, an $M$-dimensional subspace $E_k \subset \mathcal{H}_\Gamma$, such that

$$\sup_{u \in E_k} \frac{(u, \mathbf{Y}_k u)}{(u, \mathbf{J}_k u)} \leq -\lambda_k, \tag{5.9}$$

and $\lambda_k \to \infty$ as $k \to \infty$. Let now $P_k$ denote the orthogonal projection on the $M$-dimensional subspace $\Lambda^{-1/2} E_k$, where $\Lambda = \left(1 + (i\partial_s)^2\right)^{1/2}$. It follows from (5.9) that

$$P_k \Lambda^{1/2} \mathbf{Y}_k \Lambda^{1/2} P_k \leq 0, \tag{5.10}$$

$$\lim_{k \uparrow k_0} P_k \Lambda^{1/2} \mathbf{J}_k \Lambda^{1/2} P_k = 0. \tag{5.11}$$

Let $Q_{\varepsilon,k}$ denote the spectral projection of $\Lambda^{1/2}\mathbf{Y}_k\Lambda^{1/2}$ corresponding to $(-\infty, \varepsilon]$. By Eq.(3.2) one can choose $\varepsilon > 0$ in such a way that this projection is finite dimensional and analytic in $k$ for $k$ near $k_0$. From Eq.(5.10), we obtain $\mathrm{Ran}(P_k) \subset \mathrm{Ran}(Q_{\varepsilon,k})$ and hence

$$Q_{\varepsilon,k} P_k Q_{\varepsilon,k} = P_k . \tag{5.12}$$

Taking a weakly converging subsequence, w-$\lim_{n\to\infty} P_{k_n} = P_\infty$, Eq.(5.12) implies that $\lim_{n\to\infty} P_{k_n} = P_\infty$ holds in fact in the norm topology. Therefore, $P_\infty$ is an orthogonal projection on an $M$-dimensional subspace of $\mathcal{H}_\Gamma$. It follows further from Eq.(5.11) that this subspace is in the kernel of $\Lambda^{1/2}\mathbf{J}_{k_0}\Lambda^{1/2}$. Therefore,

$$\dim \ker\bigl(\mathbf{J}_{k_0}|_{\mathcal{H}_\Gamma^{-1/2}}\bigr) \geq M .$$

We complete the proof by using Lemma 4.2, which implies

$$\dim \ker\bigl(\mathbf{A}_{k_0}\bigr) \geq M .$$

Thus, by Lemma 4.5, there are at least $M$ eigenvectors of $-\Delta_\Omega$ with eigenvalue $k_0^2$. The proof of the second half of the Main Theorem is complete. $\square$

**Sketch of the connection between Eq.(2.15) and (2.16).** We have shown Eq.(2.15) in (4.14). Since we assume that $k^2 \notin \sigma(-\Delta_\Omega)$, we can write $\mathcal{L}_k = U_k|\mathcal{L}_k|$, with $U_k$ unitary, by Lemma 5.3. Therefore, for any $u \in \mathcal{H}_\Gamma$, we have, omitting the subscript $k$, and with scalar products in $\mathcal{H}_\Gamma$,

$$\begin{aligned}
(Uu, \mathbf{S}_k Uu) &= (Uu, Uu) - 2\pi i(Uu, \mathcal{L}\mathbf{A}^{-1}\mathcal{L}^* Uu) \\
&= (u, u) - 2\pi i(\mathcal{L}^* Uu, \mathbf{A}^{-1}\mathcal{L}^* Uu) \\
&= (u, u) - 2\pi i(|\mathcal{L}|u, \mathbf{A}^{-1}|\mathcal{L}|u) \\
&= (u, u) - 2i(\mathbf{J}^{1/2}u, \mathbf{A}^{-1}\mathbf{J}^{1/2}u) \\
&= (v, \mathbf{J}^{-1}v) - 2i(v, \mathbf{A}^{-1}v) \\
&= (\mathbf{A}w, \mathbf{J}^{-1}\mathbf{A}w) - 2i(\mathbf{A}w, w) \\
&= (\mathbf{A}w, \mathbf{J}^{-1}(\mathbf{A}w - 2i\mathbf{J}w)) \\
&= (\mathbf{A}w, \mathbf{J}^{-1}(\mathbf{Y}w - i\mathbf{J}w)) \\
&= (\mathbf{A}w, \mathbf{J}^{-1}\mathbf{A}^* w) = (v, \mathbf{J}^{-1}\mathbf{A}^* w) \\
&= (u, \mathbf{J}^{-1/2}\mathbf{A}^*\mathbf{A}^{-1}\mathbf{J}^{1/2}u) \\
&= (u, \mathbf{J}^{-1/2}\widetilde{\mathbf{S}}\mathbf{J}^{1/2}u) ,
\end{aligned}$$

where $v = \mathbf{J}^{1/2}u$, and $w = \mathbf{A}^{-1}v$.

**Acknowledgments.** 

profited from numerous discussions with him, which helped us to sharpen our outlook on this nice problem. A useful discussion with Peter Buser has led to the examples of Section 1. Our work was completed in the nice atmosphere of the Weizmann Institute, with support from the Fonds National Suisse, a Julius Baer Fellowship at the Weizmann Institute for JPE, and the Einstein Center of the Weizmann Institute.